\documentclass[aps,
 prx,
 reprint,
superscriptaddress,
nofootinbib,
nobibnotes,
amsmath,amssymb,
nolongbibliography,
noeprint,
floatfix
]{revtex4-2}
\usepackage{caption}
\usepackage{soul}
\usepackage{ragged2e}  
\usepackage{float}
\usepackage{graphicx}
\usepackage{mathrsfs}  
\usepackage{amsthm}
\usepackage{amssymb}  
\usepackage{bm} 
\usepackage{physics} 
\usepackage[pdftex,colorlinks=true,linkcolor=blue,citecolor=blue,urlcolor=blue]{hyperref}

\usepackage{tikz}
\usepackage[utf8]{inputenc}
\makeatletter
\renewcommand{\@makecaption}[2]{%
  \vskip\abovecaptionskip
  \begin{minipage}{\linewidth} % Force full width
  \small
  \parbox{\linewidth}{\justify #1. #2} % Justify the caption text
  \end{minipage}
  \vskip\belowcaptionskip
}
\makeatother
\usepackage{placeins}
\usepackage{hyperref}% add hypertext capabilities
\begin{document}

\preprint{APS/123-QED}
\title{Relativistic Tidal Dissipation and the Gravitational-Wave Signal\\
of a White Dwarf Orbiting an Intermediate-Mass Black Hole}

\author{Yang Yang}
\affiliation{Department of Astronomy, School of Physics, Peking University, 100871 Beijing, China}
\affiliation{Kavli Institute for Astronomy and Astrophysics, Peking University, 100871 Beijing, China}

\author{Leif Lui}
\affiliation{Beijing Institute of Mathematical Sciences and Applications, Beijing 101408, China}

\author{Alejandro Torres-Orjuela}
\affiliation{Beijing Institute of Mathematical Sciences and Applications, Beijing 101408, China}

\author{Xian Chen}
\email[Contact author: ]{xian.chen@pku.edu.cn}
\affiliation{Department of Astronomy, School of Physics, Peking University, 100871 Beijing, China}
\affiliation{Kavli Institute for Astronomy and Astrophysics, Peking University, 100871 Beijing, China}

\date{\today}

\begin{abstract}
Finding intermediate-mass black holes (IMBHs) and measuring their masses and spins are key to understanding massive black hole formation.  
White dwarf (WD)–IMBH binaries provide a unique probe because they emit both electromagnetic radiation and gravitational waves (GWs), thereby conveying richer information. 
However, such multi-messenger sources often enter the regime of strong gravity, where existing models often fail to capture their relativistic dynamics. 
Here, we develop a fully relativistic model for the tidal response of a WD close to an IMBH, and use it to study the secular orbital evolution as well as the GW signal.
We find that for IMBHs more massive than $10^5\,M_\odot$, tidal interaction becomes relativistic and sensitive to IMBH spin.
The interaction generally dissipates binary orbital energy and angular momentum, but due to relativistic frame rotation, which reduces phase coherence across pericenter passages, the orbit-averaged tidal dissipation rate can be suppressed by up to $\sim 50\%$ relative to Newtonian predictions. Including tidal dissipation leads to more rapid damping of the orbital eccentricity, to the extent that the pericenter distance may even increase over time, potentially explaining quasi-periodic eruptions and secular orbital period growth. Such tidal effect accumulates into measurable phase and amplitude deviations in the GW signal. 
For typical space-based observations, the GW waveform
mismatch can reach $\mathcal{O}(0.1)$ within 6 months.  
Our results indicate that relativistic tidal dissipation is both dynamically important and observationally essential for reliably predicting the multi-messenger signals of WD-IMBH systems.
\end{abstract}


\maketitle
\section{Introduction}

Intermediate-mass black holes (IMBHs), with masses in the range of $ M_\bullet
\sim 10^2-10^{5} M_{\odot} $, fill the intriguing gap between stellar-mass
black holes (BHs) and supermassive BHs.  
Their mass and spin in principle
encode unique information about how the most massive BHs in the universe come into being~\cite{Volonteri2010,Latif_Ferrara_2016,Inayoshi_2020}.
Their spatial and redshift distributions are also expected to hold vital clues about the role massive BHs play in shaping their host
galaxies~\cite{Kormendy_2013,Silk2017,Greene2020}. 
However, despite many efforts to find their electromagnetic (EM) signatures using conventional astronomical methods, observational evidence for IMBHs remains scarce~\cite{Reines_2016,Mezcua2017, Greene2020}.

The detection of gravitational waves (GWs) in the past decade by ground-based
interferometers~\cite{LIGO_O3, LIGO_O4_2026} and pulsar timing
arrays~\cite{Reardon_2023, Xu_2023, EPTA_2023, Miles_2024, NanoGRAV_2025} opens
up new possibilities of finding IMBHs.  It has been predicted that in
dense stellar environments such as globular clusters, IMBHs can capture compact
objects, such as stellar-mass BHs or other IMBHs, and a fraction of such binaries can coalesce via GW radiation within a Hubble time \cite{Hils1995, Baumgardt_2004,
Blecha_2006, Gurkan_2006, Amaro_Seoane_2006}.  During the last few months of
evolution, these binaries emit GWs in the frequency band of $10^{-3}-0.1$ Hz, which
is the optimal observational band for space GW missions such as LISA, Taiji,
TianQin, and lunar-based detectors~\cite{Luo_2016, Luo_2021, Amaro_Seoane_2023, Ajith_2025}.

Among the systems that may harbor IMBHs, a binary composed of an IMBH and a
white dwarf (WD) stands out as a unique multi-messenger target because it emits
both EM and GW signals during its last evolutionary stage
\cite{Ivanov2007,Menou2008,Sesana2008,Zalamea2010}.  On one hand, before the WD
falls beyond the event horizon of the IMBH, mass transfer (MT) or tidal
disruption can happen \cite{Luminet_1989, Rosswog2009, MacLeod2014, Chen_2023}.
They produce EM counterparts. On the other hand, thanks to the relatively high
density of WD, tidal disruption events (TDEs) or MT occur not too far from the event horizon
of the IMBH, creating a strong-field dynamical condition for powerful GW
emission~\cite{Hils_1995, Sesana2008,Zalamea2010}. Therefore, a joint detection
of EM and GW signals  would unambiguously place the (otherwise hidden) BH in
the intermediate-mass range and significantly improve parameter estimation by
breaking degeneracies inherent to GW-only observations~\citep{Kocsis_2011,
Pfister_2022, Chen_2023, Kejriwal_2024}.

We may have already witnessed the EM outbursts of these relativistic WD-IMBH
binaries.  It has been suggested that the population of X-ray transients known
as quasi-periodic eruptions (QPEs) are possibly such mass-transferring systems
whose orbital evolution is driven mainly by GW radiation ~\cite{Dai_2009, Dai2011, Xian_2021, Wang_2022,
King2022, Chen_2022, ZhaoZ2022, Suzuguchi_2025, Lui_2025}. More speculative evidence may
come from rare high-energy transients.  For example, a recently reported
peculiar long-duration gamma-ray burst has been proposed to originate from the
tidal disruption of a WD by an  IMBH~\cite{Neights_GCN_2025, LiDongyue_2025,
Levan_2025, Eyles-Ferris_2026}.  If these EM events are associated with
WD–IMBH systems, they form a useful sample that can be used to verify the
detectability of IMBHs by future GW observations.

To extract useful information from these potential EM counterparts, a
self-consistent dynamical model for WD–IMBH interaction is required.  Most
existing models assume that the orbital evolution is driven primarily by GW
radiation, while the EM emission is powered by the mass stripped from the WD
and accreted onto the IMBH
\cite{Sesana2008,Zalamea2010,King_2020,King2022,Wang_2022,Chen_2023,
WangDi2024,YangYang2025}.  However, compared to other compact objects such as
neutron stars or stellar-mass BHs, the main difference of a WD lies in its
relatively large size.  Therefore, a test particle, as has been assumed in the
aforementioned models, is insufficient to capture the rich dynamics induced by
the tidal response of the WD.  Indeed, models that include tidal heating of
the WD and dissipation of the orbital energy do show results that deviate from
pure test-particle evolution
\cite{Lai_1997,Ivanov2007,Vick2017,Wang_2022,Lau2025}.

Despite the importance of tidal dissipation (TD), when dealing with the tidal force imposed on the WD, existing models commonly adopt the  Newtonian prescription (e.g., \cite{Vick2017,Wang_2022,Lau2025}), even though the system may have already entered the regime of strong-gravity.   
For our interest, producing a GW signal with a high signal-to-noise ratio (SNR) requires the pericenter $r_{p}$ of the WD-IMBH binary to lie within a few tens of the IMBH’s gravitational radius
$r_{g}$~\cite{Amaro_Seoane_2007,Chen_2023}. Therefore, we need a relativistic version of the TD model.
 
In a few exceptions, the relativistic tidal force is evaluated in a Fermi
normal coordinate (FNC) frame that follows the motion of the
WD~\cite{Marck1983,Ishii2005}.  Hydrodynamical simulations performed in the FNC
frame further indicates that the energy transfer during pericenter passage can
differ substantially from the Newtonian prediction~\cite{Cheng2013}.  Nevertheless,
owing to computational limitations, these simulations so far have not been able
to follow the WD through multiple pericenter passages.  The
cumulative impact on the WD and the long-term evolution of the binary remains
unclear, and hence the theoretical templates
for joint EM and GW analyses still lack the necessary fidelity required to guide
observations and extract reliable physical parameters.

To overcome this obstacle, we develop a model for calculating the tidal energy and angular-momentum fluxes of a WD in the strong-gravity regime near an IMBH. 
We formulate the problem in FNCs, which provide a locally inertial frame along the stellar orbit and allow the tidal interaction to be treated using Newtonian stellar hydrodynamics even though the spacetime background is essentially curved. 
TD is modeled through propagating gravity-wave dynamical tides under outgoing-wave boundary conditions~\cite{Fuller2012,Vick2017}, motivated by the distinctive density structure of WDs.

The paper is organized as follows. We first outline the physical framework, quantitatively explaining the necessity of relativistic tidal fields, and summarize the key assumptions of our relativistic tidal model in Sec.~\ref{sec:2}.  
In Sec.~\ref{sec:3}, we develop a relativistic treatment of TD for highly eccentric WD-IMBH binaries and describe how the resulting energy and angular-momentum fluxes are incorporated into the orbital evolution.  
We then present the resulting orbital evolution of the system in Sec.~\ref{orbital_evolution_section}, where we also take into account the effects of IMBH spin.  
In Sec.~\ref{sec:5}, we compute the GW
emission including corrections due to relativistic TD, and evaluate the detectability
of the TD effects in future GW observations.
In Sec.~\ref{sec:6}, we discuss the implications of our results and outline
potential directions for future extensions of this work.

\section{Characteristic Scales}  
\label{sec:2}

\subsection{Domain of Relativistic Correction and Quadrupole Tide}
\label{sec:2-1} 

\begin{figure}   % h=here, t=top, b=bottom, p=page
    \centering
    \includegraphics[width=\linewidth]{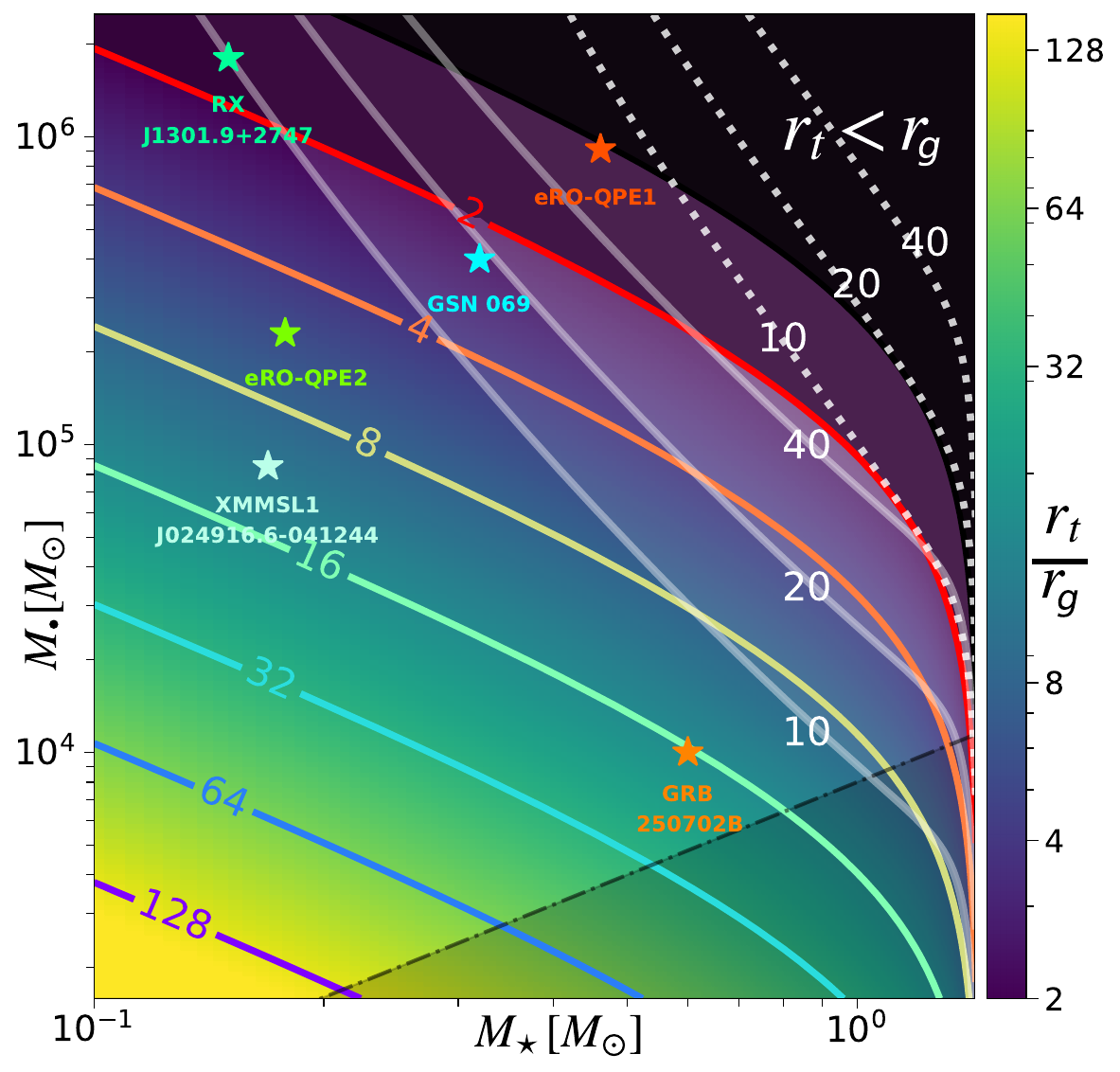}  
    \caption{
    Ratio of the Newtonian TDE radius of a WD, $r_{t}$, to the gravitational radius of an IMBH, $r_{g}$, 
	shown in the $M_{\star}$-$M_{\bullet}$ parameter space. 
    The white lines indicate constant  SNRs  ($=10$, $20$, $40$) for LISA observation, assuming a pericenter distance of $r_p = 2\,r_t$, a redshift of $z=0.02$ ($d_{L} \simeq  85 $Mpc), and an observation period of $4$ years, whereas the solid and dotted line styles correspond to orbital eccentricities of $e=0.95$, $0.99$, respectively. 
    The region below the black dot-dashed line in the lower-right corner corresponds to  $R_{\star}/r_{t} >  5\%$, where the assumption of quadrupole tidal field becomes invalid. The stars mark the inferred locations of a sample of
	QPEs.
}
    \label{fig:rt_rg_contour_plot}  
\end{figure}

The MT and TD processes of a WD around IMBH are often
affected by strong general-relativistic (GR)
effects~\cite{Ivanov2007,Menou2008,Sesana2008,Chen_2023}.
To identify the regime relevant to relativistic correction, we show in FIG.~\ref{fig:rt_rg_contour_plot} the ratio between the tidal disruption radius
(TDE radius), $r_t\,$$=$$\,R_{\star}(M_{\bullet } /M_{\star})^{1/3}$, and the
gravitational radius, $r_g=GM_{\bullet }/c^2$, in the 
$M_{\star}$-$M_{\bullet}$ parameter space.  Here $M_{\bullet}$ denotes the mass of the
IMBH, while $M_{\star}$ and $R_{\star}$ are the mass and radius of the  WD,
respectively.  The figure also shows the inferred locations of several observed
QPEs \cite{Miniutti2019, Giustini2020, Arcodia2021,
Chakraborty2021,Neights_GCN_2025}, whose physical parameters were obtained by
fitting their light curves using the mass-transfer models, assuming highly
orbital eccentricities \cite{Chen_2022, Eyles-Ferris_2026}. 

We find that when the IMBH mass exceeds $\sim10^{5}\,M_\odot$, the TDE radius
will satisfy $r_t\lesssim10\,r_g$, indicating strong relativistic effects 
during the processes of tidal heating, mass transfer, or tidal disruption. 
In the same parameter
space where $r_t/r_g\lesssim10$, if the WD is more massive than $0.5M_{\odot}$, such as a
carbon–oxygen (CO) WD~\cite{Benvenuto1999}, the SNR of the GW radiation in the
LISA band during an observational period of four years will exceed $10$, as is
indicated in FIG.~\ref{fig:rt_rg_contour_plot} by the solid and dotted white
curves (calculated following Refs.~\cite{Finn2000, Barack2004}).  Therefore,
high-SNR GW events are highly relativistic as well. These findings imply
that multi-messenger WD-IMBH systems discovered in the future by joint EM and
GW observations will display significant GR effects.

\begin{figure}   % h=here, t=top, b=bottom, p=page
    \centering
    \includegraphics[width=\linewidth]{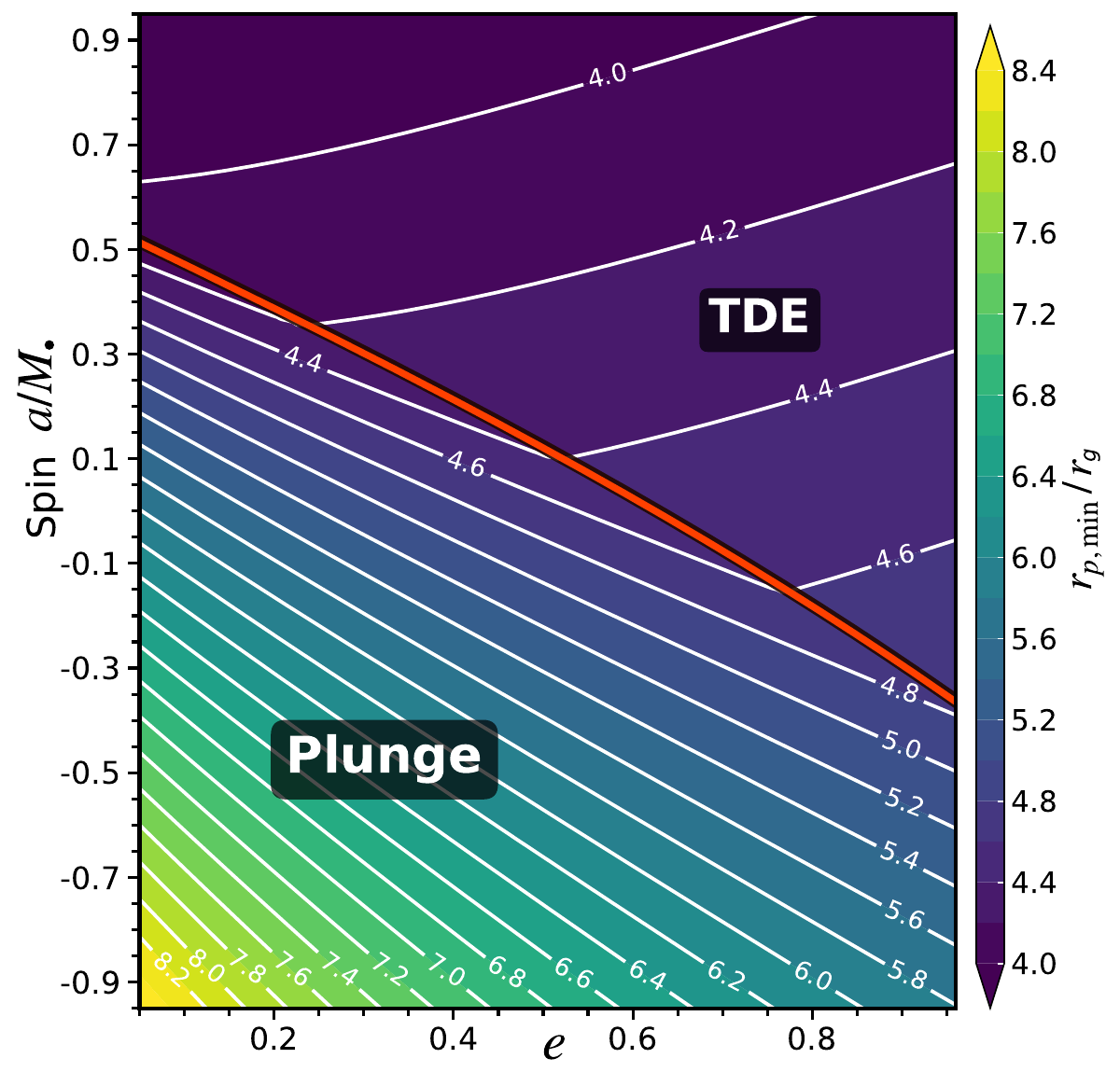}
    \caption{
    Minimum allowed pericenter distance for a WD–IMBH system due to either relativistic plunge or tidal disruption,
shown in the parameter space of IMBH spin $a$ and orbital eccentricity $e$.
   The red line marks the points at which the critical plunge radius $r_{\rm plunge}(a,e)$
	equals the critical TDE radius $r_{\rm TDE}(M_{\star},M_{\bullet},a,e)$. Therefore, in the
    lower-left region, $r_{\rm plunge}>r_{\rm TDE}$, and a WD will plunge into IMBH
before tidal disruption. On the contrary, a WD in the upper-right region gets tidally disrupted before plunging.
    Here we have assumed $(M_{\star},M_{\bullet}) = (0.6\,M_{\odot},10^{5}\,M_{\odot})$.
}
    \label{fig:TDE_Plunge_limit}  
\end{figure}

The presence of GR effects complicates the calculation of the tidal force
exerted by IMBHs on WDs.  To approximate the tidal field, previous works often
decompose the field around the WD into multipoles, starting from the quadrupole
moment~\cite{Press1977, Lai_1997, Fuller2012, Ivanov2004, Will2016, Vick2017},
and the relative contribution between adjacent multipoles scales with
$R_{\star}/r_p \sim R_{\star}/r_t \sim (M_{\star}/M_{\bullet})^{1/3}$.  For WDs
around an IMBH of $M_{\bullet} \gtrsim 10^{4}\, M_{\odot}$, this ratio remains
below $5\%$, indicating the predominance of the quadrupole moment.  In
FIG.~\ref{fig:rt_rg_contour_plot}, the black dot-dashed line corresponds to
$R_{\star}/r_p=5\%$. Above this line is the parameter space where the tidal
field can be well approximated by a quadrupole. The majority of our interested
systems lie in this region; therefore, we will derive only the quadrupole tidal
field in the following.

\subsection{Plunge Versus TDE}
\label{sec:2-2}

For the purposes of modeling a WD-IMBH system as well as predicting the
observables, it is important to understand how the evolution terminates.
Normally, the system ends up in a tidal disruption of the WD, or the WD
plunging into the IMBH. 
In the Newtonian case, the outcome is determined by the
masses of the WD and the IMBH~\cite{Hills1975,Rees1988,Sesana2008,Chen_2023}. 
In the relativistic regime, however, the orbital parameters of the WD and the spin $a$
of the IMBH\footnote{Note that $a$ denotes the Kerr spin parameter and should
not be confused with the orbital semi-major axis.} also matter,
because they determine the existence and location of the
inner-most turning point of a bound geodesic
\cite{Bardeen1972,Glampedakis2002,Levin2009,Misner2018,Stein2020}, as well as
the tidal field felt by the WD in its rest frame~\cite{Marck1983,Cheng2013}. 

To simplify the problem, in the following we will restrict the WD orbit in the
equatorial plane of the spinning IMBH. In this case, given the spin $a$, a secular orbit
admits two radial turning points, corresponding to the pericenter and
apocenter, with radial coordinates $ r_p$ and $r_a$ \cite{Bardeen1972}. Then,
following Ref.~\cite{Glampedakis2002}, we can characterize such a relativistic
orbit with two parameters, $r_p$ as is defined above and the eccentricity $e
\equiv (r_a-r_p)/(r_a+r_p)$.  

With this, we can now determine the outcome of a relativistic WD-IMBH inspiral.  This is done by
fixing the orbital eccentricity $e$ and calculating the critical pericenter distances, 
$r_{\rm plunge}(a,e)$ and $r_{\rm TDE}(M_{\star},M_{\bullet},a,e)$, below which plunge and TDE,
respectively, will occur. The larger one of these two pericenter distances determines
the actual fate of the system, setting the minimum allowed distance between the
WD and the IMBH to be
\begin{equation}
    r_{p,\mathrm{min}}(M_{\star},M_{\bullet},a,e)
	= \max\!\left(r_{\rm plunge},\, r_{\rm TDE}\right).
    \label{rpmin}
\end{equation}
Here, we have used a relativistic tidal tensor
\cite{Kesden2012,Rossi2021, xin2025GRTidal} to find $r_{\rm TDE}$, and more
details are given in Section~\ref{sec:3}.  Notice the different between $r_{\rm
TDE}$ and $r_t$, the latter of which does not depend on $a$ or $e$.

FIG.~\ref{fig:TDE_Plunge_limit} shows the minimum pericenter distance defined
by Eq.~(\ref{rpmin}) in the $e-a$ parameter space, where negative values of the
spin $a$ corresponds to retrograde orbits. This shows the impact of $a$
and $e$ on the outcome, even when the masses, $(M_{\star}, M_{\bullet})$, are both fixed.  

In particular, given $e$, we find
$r_{p,\mathrm{min}}$ decreases with increasing BH spin in both the plunge
region and the TDE region.  This anti-correlation reflects how BH spin reshapes both
the angular-momentum barrier of equatorial geodesics and the strength of
relativistic tidal field.  For prograde orbits, frame dragging lowers the
effective angular-momentum barrier, allowing a given pericenter distance to be
reached with smaller specific angular momentum and orbital energy, thereby
shifting the plunge separatrix inward~\cite{Bardeen1972,Glampedakis2002}.
Simultaneously, the anti-correlation between the relativistic tidal tensor and
the spin parameter weakens the tidal stretching at fixed $r_p$, so that tidal
disruption requires deeper penetration into the gravitational potential.  For
retrograde orbits, both effects act in the opposite direction, shifting the
plunge and TDE thresholds outward.

The contour spacing in FIG.~\ref{fig:TDE_Plunge_limit} also suggests that as
the IMBH spin increases, angular-momentum barrier shrinks much faster than the
relativistic tidal tensor weakens.  As a result, increasing $a$ will eventually
put the WD-IMBH in the TDE regime.  For this reason, for a $0.6M_\odot$ WD
around a $10^5M_\odot$ IMBH, the outcome is always a TDE when
$a\gtrsim0.5\,M_{\bullet}$.  On the contrary, when
$a\lesssim-0.3\,M_{\bullet}$, the result is plunge.

\subsection{Gravitational Radiation Versus \\
Tidal Dissipation}
\label{sec:2-3}

From the entry into the relativistic regime until its
termination by tidal disruption or plunge, the evolution of a WD-IMBH binary is driven mainly by GW radiation and tidal interaction.  
While GW radiation extracts energy and angular momentum from the binary orbit~\citep{Peters_1963, Peters_1964}, tidal interaction leads to excitation and damping of stellar oscillation modes, nonlinear tidal responses, or shock heating in the outer layers of the WD \citep{Fuller2012,YuHang2020}, thereby irreversibly dissipating orbital energy and angular momentum into internal motions, heating, and spin of the WD \citep{Press1977}.
The relative importance of GW radiation and TD can be evaluated by their respective characteristic timescales~\citep{Vick2017}.

For example, GW radiation extracts orbital angular momentum $J_{\mathrm{orb}}$ at a rate of
\begin{equation}\label{eq:tGW}
t_{\mathrm{GW}}
    = \frac{J_{\mathrm{orb}}}{\left|\dot{J}_{\mathrm{GW}}\right|}
    = \frac{5}{32}\frac{G M_{\bullet} }{c^{3}}
    \frac{q^{5/3} k^{4}}{\mathcal{C}_{\star}^{4}(1-e)^{3/2}}
    \frac{(1+e)^{5/2}}{(1+7e^{2}/8)},
\end{equation}
where $q = M_{\star}/M_{\bullet} \sim 10^{-6}$ is the mass ratio of the binary, $\mathcal{C}_{\star} = G M_{\star}/ R_{\star}c^{2} \sim 10^{-4}$ is the compactness of WD, and $k = r_p/r_t$ denotes the ratio of pericenter distance to the Newtonian tidal disruption radius.   
And the characteristic timescale associated with TD is 
\begin{equation}
t_{\mathrm{tide}}
    = \frac{J_{\mathrm{orb}}}{\left|\dot{J}_{\mathrm{tide}}\right|}
    = \frac{G M_{\bullet} }{c^{3}}
    \frac{q^{1/3} k^{13/2}}{\mathcal{C}_{\star}^{3/2}\,\dot{\mathcal{J}}\, (1-e)^{3/2}}
    (1+e)^{5/2}
\end{equation}
 \cite{Vick2017}, where $\dot{\mathcal{J}} \sim 10^{-6}$ for $k \sim \mathcal{O}(10)$ is the dimensionless tidal angular momentum transfer rate $\dot{\mathcal{J}}(k,e)$.

The ratio of the two timescales for high eccentricity is therefore
\begin{equation}
    \frac{t_{\mathrm{GW}}}{t_{\mathrm{tide}}}
    = 
    \frac{q^{4/3}\,\dot{\mathcal{J}}}{\mathcal{C}_{\star}^{5/2} k^{5/2}}
    \frac{5/32}{(1+7e^{2}/8)}\propto  \dot{\mathcal{J}}(k) k^{-5/2}.
\end{equation}
Here, using orbital energy rather than angular momentum leads to the same conclusion.    
In our problem, since both $q$ and $\dot{\mathcal{J}}$ are extremely small, significant TD appears at $r_{p}/r_{g} \lesssim 20$, corresponding to $k \lesssim 6$. 
This estimation indicates that a WD in the most relativistic
regime of an IMBH is also experiencing strong TD.

As $r_{p}$ gradually approaches $r_{t}$, the tidal interaction near pericenter becomes stronger, and the characteristic tidal forcing frequency increases, corresponding to a more rapid temporal variation of the tidal field. 
In this regime, the rapid variation of the tidal force does not allow the fluid to maintain instantaneous equilibrium. 
The tidal response can no longer be treated as a quasi-static perturbation.
To understand how the rapidly varying tidal interaction near pericenter excites dissipation inside the WD, we further analyze the characteristic tidal forcing frequency that dominates this process.  
The characteristic forcing frequency at pericenter can be estimated as
\begin{equation}\label{eq: omega_p}
    \omega_{p} = \omega_{\mathrm{orb}} \sqrt{\frac{1 + e}{(1 - e)^{3}}}
= \omega_{\star} \sqrt{\frac{1 + e}{k^{3}}},  
\end{equation}
where $\omega_{\star}$\,$= $\,$\sqrt{G M_{\star}/R_{\star}^{\,3}}$ is the stellar dynamical frequency and $\omega_{\mathrm{orb}}$\,$=$\,$ \omega_{\star} \left[ (1 - e)/k \right]^{3/2} $ is the mean orbital frequency. 
According to Eq.~\eqref{eq: omega_p}, for encounters with $k\sim 3 - 6$, the characteristic tidal forcing frequency is typically about $\sim 0.1-0.5$ of the eigenfrequency of the WD. 
In this frequency range, dynamical effects associated with the driving frequency, such as resonant excitation and phase evolution, become important.
Within this framework, the time-dependent tidal force can effectively couple to the low-order discrete eigenmode spectrum of the WDs~\cite{Lai_1997,Fuller2012,Vick2017,MacLeod2014,Lau2025}.

However,  the tidal response in this frequency range is not limited to the excitation of discrete eigenmodes. 
Gravity-wave dissipation can remain efficient, even when the tidal forcing frequency is only slightly below the stellar dynamical frequency.
This arises from the unique internal structure of WDs, which exhibit strong stratification and sharp composition gradients at the C/He and He/H transition layers~\cite{Koester1990,Althaus2010}.
These composition gradients produce strong spatial variations in the Brunt--V\"{a}is\"{a}l\"{a} frequency, which sets the maximum frequency for internal gravity wave propagation. 
As a result, the coupling between high-frequency gravity-wave structures and the tidal forcing is enhanced, allowing efficient excitation of outgoing gravity waves even in the relatively high-frequency regime.
As these gravity waves propagate toward the low-density outer envelope, conservation of wave energy flux causes their amplitudes to grow substantially, eventually leading to nonlinear wave breaking near the stellar surface.

Although low-order global oscillations may also contribute when the tidal forcing frequency becomes a non-negligible fraction of the WD dynamical frequency, the outgoing-wave dynamical tide provides a practical framework for modeling the secular tidal energy and angular-momentum transfer in highly eccentric WD--IMBH systems.  
In this work, we adopt the outgoing-wave dynamical tide formalism developed by~\citet{Fuller2012}, in which the tidal response is modeled through propagating gravity waves rather than discrete standing-wave $g$-mode resonances.

\section{Relativistic Tidal Dissipation}
\label{sec:3}

\subsection{Relativistic Tidal Field}
\label{sec:3-1}

The formulation of a TD theory in the strong-gravity regime must begin with a
rigorous relativistic treatment of the tidal force, which we briefly review in this
subsection.  The most natural coordinate system in which this force can be expressed 
is a local, freely-falling
reference frame comoving with the center-of-mass of the WD.  In
fact, this is how the FNCs are
introduced~\cite{Manasse1963,Marck1983,Ishii2005}. 

To understand the definition of FNCs, we first describe the spacetime around a spinning IMBH. In 
Boyer–Lindquist coordinates (BLCs) $x^{\mu}_{\mathrm{BLC}} = (t, r, \theta, \phi)$ and using
geometrized units, the metric takes the form
\begin{equation}\label{eq:Kerr_metric}
\begin{aligned}
\dd s^2=&-\left(1-\frac{2M_{\bullet}r}{\Sigma}\right)\dd t^2
-\frac{4M_{\bullet}ar\sin^2\theta}{\Sigma}\dd t\dd\phi
+\frac{\Sigma}{\Delta}\dd r^2\\
&+\Sigma\,\dd\theta^2
+\frac{(r^2+a^2)^2-\Delta a^2\sin^2\theta}{\Sigma}\sin^2\theta\dd\phi^2
\end{aligned} 
\end{equation}
\cite{Kerr_1963}, 
where $\Sigma=r^2+a^2\cos^2\theta$ and $\Delta=r^2+a^2-2M_{\bullet}r$. Notice
that geodesic motion in this spacetime admits three constants of motion $(E,
L_{z}, Q)$, which are the specific energy, $z$-angular momentum, and Carter
constant, respectively \cite{Carter_1968}. They will be used to construct
the FNCs and the corresponding tidal tensor.

To relate vectors and tensors in different coordinate systems, orthonormal tetrads are
often introduced.  In BLCs, the tetrad bases are $\boldsymbol{e}^{(a)}=
e^{(a)}_{\mu}\,\mathbf{d}x^{\mu}_{\mathrm{BLC}}$, while in FNCs the
corresponding bases are $\mathbf{\Lambda}^{(a)}=
\Lambda^{(a)}_{\alpha}\,\mathbf{d}x^{\alpha}_{\mathrm{FNC}}$.  In both cases,
$(a)=0,1,2,3$ are the tetrad indices and $\mathbf{d}x^{\mu}$ denotes the
coordinate one-form.  We further denote the FNC coordinates as
\(x^{\alpha}_{\mathrm{FNC}} = (\tau, x^{1}, x^{2}, x^{3})\) to distinguish them
from the BLCs, and hence \(\tau\) is the proper time along the reference
geodesic.  Explicit expressions for the tetrad components $e^{(a)}_{\mu}$ and
$\Lambda^{(a)}_{\alpha}$ can be found in Refs.~\cite{Marck1983,Ishii2005}.
Here, we emphasize the relation
$\partial x^{\mu}_{\mathrm{BLC}} / \partial x^{\alpha}_{\mathrm{FNC}}
= \Lambda^{(a)}_{\alpha} e^{\mu}_{(a)}$.
Based on it, the transformation of tensor in different coordinates can be
performed. 
For example, the Riemann tensor transforms as
\begin{equation}
    R^{\mathrm{FNC}}_{\alpha\beta\gamma\delta}
    =
    \Lambda^{(a)}_{\alpha}\, \Lambda^{(b)}_{\beta}
    \Lambda^{(c)}_{\gamma}\, \Lambda^{(d)}_{\delta}
    e^{\mu}_{(a)}\, e^{\nu}_{(b)}
    e^{\kappa}_{(c)}\, e^{\lambda}_{(d)}
    R^{\mathrm{BLC}}_{\mu\nu\kappa\lambda},
\end{equation}
where $R^{\mathrm{BLC}}_{\mu\nu\kappa\lambda}$ is determined by the Boyer-Lindquist metric given
in Eq.~(\ref{eq:Kerr_metric}).

From the Riemann tensor in the FNCs, one can derive the tidal tensor. 
In the following analysis, we will only keep the leading order terms.  
Higher-order tidal corrections arise from higher-order terms in the FNC metric expansion, and involve covariant derivatives of the curvature tensor as well as nonlinear curvature couplings~\cite{Ishii2005}, and hence they can be neglected for the reasons discussed in Sec.~\ref{sec:2-1}.

In addition to higher-order tidal corrections in the FNCs, we further neglect higher-order deviations from geodesic motion induced by the finite WD spin, corresponding to spin–curvature coupling effects~\cite{Papapetrou1951,Dixon1970,Tod1976,Mortazavimanesh2009,Drummond_2022,Drummond_2022b,Skoupy_2025}.  
Spin-induced spacetime effects arise from momentum currents that generate the off-diagonal components of the stress--energy tensor responsible for frame dragging, so their magnitude is controlled by the relativistic nature of the mass currents.   
These currents are largest at the stellar surface where the rotational velocity peaks.  
The resulting spin--curvature force relative to geodesic acceleration scales as
$
a_{\text{spin}} / a_{\text{geo}} \sim \left(R_{\star} / r_p\right)\left(\omega_{\star} R_{\star} / c\right)
$.
However, for a WD, even if the rotation approaches the centrifugal breakup limit, the surface velocity remains non-relativistic, with $\omega_\star R_\star/c = \mathcal{C}_\star^{1/2} \sim 10^{-2}$, as a consequence of its low compactness.  
Therefore, the geodesic motion in the background Kerr spacetime  provides an accurate reference trajectory along which the tidal field is evaluated in the FNC framework, forming the basis for the TD modelling.

In modeling TD within the FNC framework, we further restrict our analysis to a
nonspinning WD on an equatorial orbit in order to isolate the baseline tidal
response, since stellar rotation modifies both the tidal forcing frequency and
the oscillation spectrum.  A self-consistent treatment of the rotational
effects would require both a prescription for spin evolution and a physically
motivated choice of spin configuration, neither of which is fully available in
the present framework.  The first limitation concerns spin evolution: the
present FNC framework does not fully characterize the torque-induced precession
of the stellar spin axis, preventing a self-consistent treatment of generic
spin orientations.  The second concerns spin configuration: most existing
treatments of TD  assume an aligned spin--orbit configuration, since misaligned
spins introduce additional dynamical complexity associated with spin-axis
evolution and complicate the mode coupling structure. This alignment, however,
may not be realistic.  Therefore, aligned spin configurations should not be
regarded as the natural baseline.  Taken together, these considerations
motivate our choice of a nonspinning WD as the baseline configuration, while a
self-consistent treatment of rotating-body tides in FNC is deferred to future
work.

With these simplifications, the relativistic quadrupole tidal tensor
$C_{ij}$ in the rest frame of the WD's center-of-mass is 
\begin{equation}\label{eq:C_tidal_tensor}
\begin{aligned}
    C_{11}=&\frac{M_{\bullet}}{r^{3}}\left[\left(1-3 \cos ^{2} \Psi\right)
    -\frac{3 \mathscr{L}}{r^{2}} \cos ^{2} \Psi\right], \\
    C_{12}=&\  C_{21}
    =-\frac{3 M_{\bullet}}{r^{3}}\left(1+\frac{\mathscr{L}}{r^{2}}\right)
    \sin \Psi \cos \Psi, \\
    C_{22}=&\frac{M_{\bullet}}{r^{3}}\left[\left(1-3 \sin ^{2} \Psi\right)
    -\frac{3 \mathscr{L}}{r^{2}} \sin ^{2} \Psi\right], \\
    C_{33}=&\frac{M_{\bullet}}{r^{3}}\left(1+\frac{3 \mathscr{L}}{r^{2}}\right),
\end{aligned}
\end{equation} 
where $\mathscr{L}=(L_z-aE)^2$. 
Notice that in this work, the FNC basis vectors labeled by indices $1$ and $2$ lie in the equatorial plane, whereas in Refs.~\cite{Marck1983,Ishii2005,Cheng2013} the equatorial plane aligns with the $1$ and $3$ directions. 
We make this modification because it allows the tidal field to be expressed more compactly in terms of spherical harmonics.

\begin{figure}   % h=here, t=top, b=bottom, p=page
    \centering
    \includegraphics[width=\linewidth]{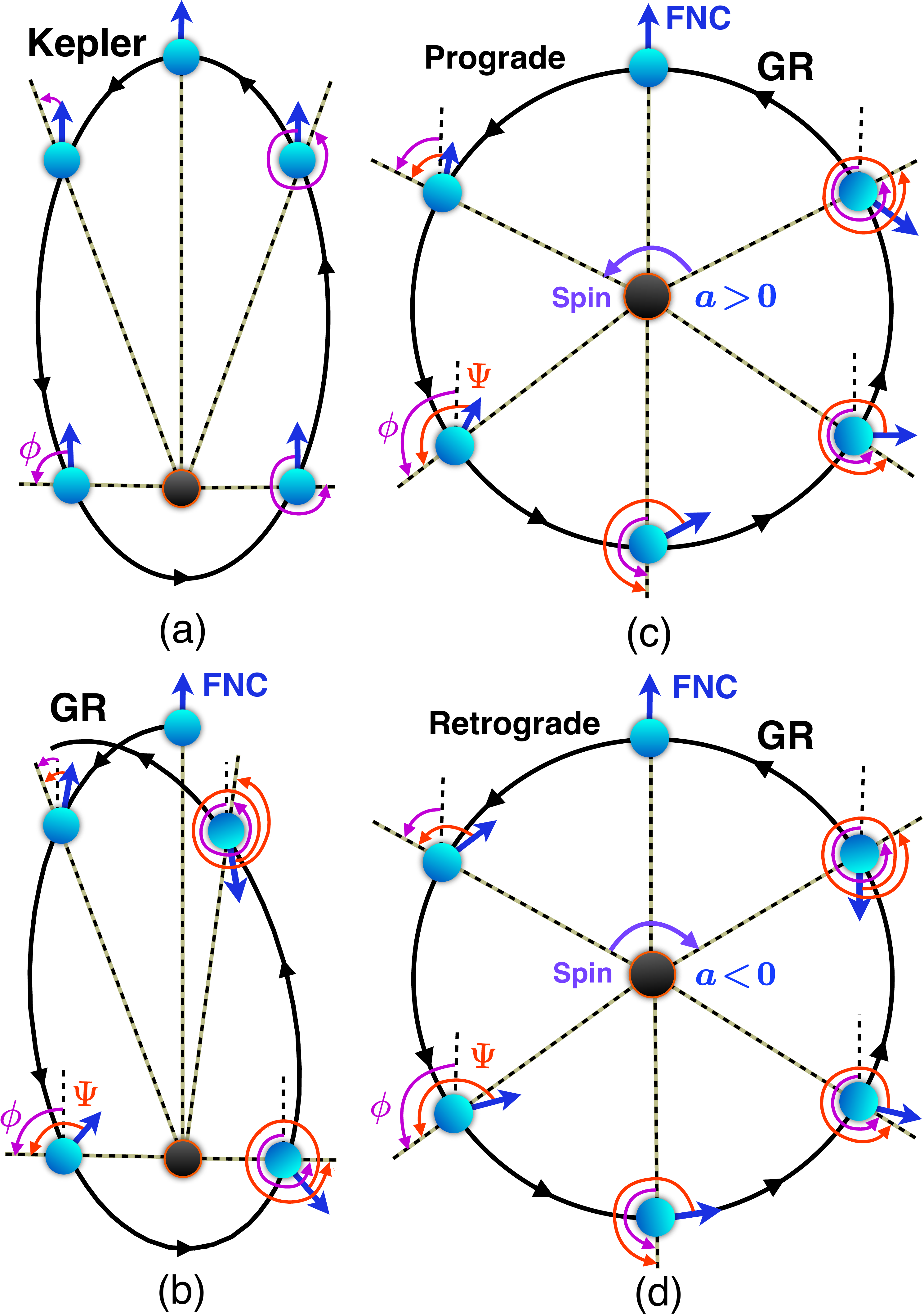}  
	\caption{Orientation of the Fermi normal coordinates for a WD on (a) an eccentric Newtonian orbit,
	(b) an eccentric orbit around a Schwarzschild IMBH, (c) a prograde circular orbit in the equatorial
	plane of a Kerr IMBH, and (d) a retrograde circular orbit in the equatorial plane of a Kerr IMBH.
	In each panel, the blue arrow represents the tetrad vector $\mathbf{\Lambda}^{(1)}$. The azimuthal angle
	$\phi$ in the Boyer-Lindquist coordinates and the rotational angle $\Psi$ associated with the 
	Fermi-normal coordinates are also marked.
}
    \label{fig:FNC_Kerr_spin_show}  
\end{figure}

The tensor components in Eq.~(\ref{eq:C_tidal_tensor}) depend explicitly on the rotation angle $\Psi$, which specifies the orientation of the parallel-transported FNC frame relative to the BLC basis \citep{Marck1983}.
Its geometrical meaning is illustrated in FIG.~\ref{fig:FNC_Kerr_spin_show}. 
In the Newtonian limit, the local inertial frame remains aligned with the instantaneous radial direction, so that $\Psi$ coincides with the orbital phase, $\Psi=\phi$ (panel a).
By contrast,  in GR, the FNC tetrad undergoes nontrivial parallel transport along the geodesic.
For eccentric geodesic motion in Schwarzschild spacetime (panel b), spacetime curvature causes the parallel-transported FNC tetrad to undergo a cumulative rotation along the orbit.
This rotation proceeds in the opposite sense to the relativistic apsidal advance of the geodesic.
When the central IMBH is spinning (panels c and d), frame dragging further modifies the evolution of $\Psi$.
The resulting rotation depends on the relative orientation of the orbital angular momentum and the BH spin: for prograde orbits the accumulated rotation of the parallel-transported FNC tetrad is partially suppressed, whereas for retrograde orbits it is enhanced.

This additional rotation occurring in the regime of strong gravity, together
with the apsidal precession of eccentric orbit, implies that the orientation of the tidal-tensor principal axes, evaluated at successive pericenter passages, is not fixed in the WD comoving frame.
This variation changes the phase of
the tidal forcing, so that it no longer repeats modulo  $2\pi$. 
Consequently, the coherence between tidal forcing and stellar
mode, which may establish in the Newtonian case, will be broken in the
relativistic regime. The efficiency of energy and angular-momentum exchange
between the WD and the binary orbit is, therefore, fundamentally altered.

It is worth mentioning that relativistic dephasing remains effective even in
the presence of stellar rotation.  Stellar rotation primarily shifts the tidal
forcing frequency, while preserving the nearly periodic structure of the tidal
forcing.  Such a frequency shift mainly modifies the resonance condition and
the effective tidal forcing frequency, but does not destroy the phase coherence
between successive orbital passages.  By contrast, the relativistic rotation of
the FNC tetrad becomes strongest near pericenter, where the orbital motion
changes the most rapidly.  This relativistic rotation prevents the tidal tensor
from repeating periodically between successive pericenter passages and causes
the tidal forcing to lose coherence over multiple orbits.  Therefore, the
suppression of TD caused by the relativitic rotation of FNC should persist
even when stellar rotation is present.

\subsection{Tidal Response of White Dwarf}\label{sec3-2}

While the construction of the tidal tensor requires a relativistic treatment, the tidal response of the WD can be described within Newtonian hydrodynamics, since the spacetime in the FNCs is locally nearly flat and the WD itself is only weakly relativistic \citep{Ishii2005,Cheng2013}.

To model this tidal response, we should specify how the tidal perturbation propagates and dissipates inside the WD.
The tidal perturbation excites gravity waves that continuously transport energy and angular momentum away from the excitation region and dissipate through nonlinear wave breaking near the stellar surface, as discussed in Sec.~\ref{sec:2-3}.
This gravity-wave behavior motivates us to adopt the outgoing-wave dynamical tide formalism.
The outgoing-wave description remains valid over most of the gravity-wave propagation region inside the WD, where the Brunt--V\"ais\"al\"a frequency — the buoyancy frequency that sets the maximum frequency for internal gravity wave propagation — satisfies $N \sim 10^{2}  \omega_\star \gg \omega_p$. 
This strong frequency hierarchy allows the gravity-wave dispersion relation used in Refs.~\citep{Fuller2012,YuHang2020} to accurately describe the local wave dynamics.

In their works, a specific tidal force $\boldsymbol{a}_{\rm tidal}$ is imposed on the WD, and the response is quantified by the Lagrangian displacement $\boldsymbol{\xi}$, expressed in spherical coordinates $(R,\vartheta,\varphi)$.
Both the displacement and the driving force are decomposed into vector spherical harmonics $\mathbf{Y}^{\mathrm{R,S,T}}_{lm}$, where the superscripts R, S, and T denote the radial, spheroidal, and toroidal basis functions, respectively \citep{Barrera_1985}.
Following this formalism, we rewrite the tidal response in the form 
\begin{equation}\label{eq:decom}
\begin{aligned}
\boldsymbol{\xi} =& \sum_{lm}
\left(
\xi^{\mathrm{R}}_{lm}\mathbf{Y}^{\mathrm{R}}_{lm}
+\xi^{\mathrm{S}}_{lm} \mathbf{Y}^{\mathrm{S}}_{lm}
+\xi^{\mathrm{T}}_{lm} \mathbf{Y}^{\mathrm{T}}_{lm}
\right),\\
\boldsymbol{a}_{\mathrm{tidal}}
=&
\sum_{lm}
\left(
F^{\mathrm{R}}_{lm} \mathbf{Y}^{\mathrm{R}}_{lm}
+F^{\mathrm{S}}_{lm} \mathbf{Y}^{\mathrm{S}}_{lm}
+F^{\mathrm{T}}_{lm} \mathbf{Y}^{\mathrm{T}}_{lm}
\right),
\end{aligned}
\end{equation} 
where the expansion coefficients depend on the radial coordinate $R$ and the proper time $\tau$.

Since the coefficients are time-dependent, they are further decomposed into Fourier series,   $F^{\mathrm{R,S,T}}_{lm}(R,\tau)=\sum_n
F^{\mathrm{R,S,T}}_{nlm}(R)e^{-i \omega_{nlm} \tau}$. 
The solution for each mode, i.e., $\xi^{\mathrm{R,S,T}}_{nlm}$, can then be obtained by solving a Schrödinger-like wave equation\footnote{
The other displacement components, $\xi^{\rm S,T}_{nlm}$, satisfy equations of similar form~\cite{Fuller2012}.  
Under the WKB approximation, the spheroidal component is related to the radial displacement by $\xi^{\rm S}_{nlm} \simeq -i \xi^{\rm
R}_{nlm}k_{nlm}R/[l(l+1)]$.  } 
\begin{equation}\label{eq: Z_second-Eq}
Z^{\prime \prime}_{nlm} + k_{nlm}^2\, Z_{nlm}
= \chi_{nlm}^{-1/2} \frac{l(l+1)}{\omega_{nlm}^2}N^2 V_{nlm}(R),
\end{equation}
where $Z_{nlm}(R):= \chi_{nlm}^{-1/2} R^2
\xi^{\rm R}_{nlm}$ is a rescaled radial displacement variable, 
$V_{nlm}(R)$ is a source term associated with the driving force, primes denote derivatives with respect to $R$, and $k_{nlm}$ is a function of $R$.

Given the decomposition in Eq.~(\ref{eq:decom}), the source term
$V_{nlm}(R)$ can be written in terms of the force components as  
\begin{equation}\label{eq:V def}
\begin{aligned}
& V_{nlm}(R)=\;
\frac{F^{\rm R}_{nlm}}{N^2}
\left(1 - \frac{\omega_{nlm}^2}{L_l^2}\right)
- \frac{ \bigl(F^{\rm S}_{nlm} R\bigr)' }{N^2}  \\
&\quad \quad + F^{\rm S}_{nlm}
\left[
\frac{R}{g}
+ \frac{1}{N^2} \frac{2\omega_{nlm}^2}{L_l^2 - \omega_{nlm}^2}
\left(
\frac{\mathrm{d}\ln c_{ s}^2}{\mathrm{d}\ln R} - 2
\right)
\right],
\end{aligned}
\end{equation}
where all background quantities,
$ g = -P^{\prime} /\rho$,
$c_s^2 = (\partial P / \partial \rho)_s$,
$N^2= g^2( \rho^{ \prime}/ P^{\prime} - 1/c_s^2)$,
and
$L_l^2= l(l+1)c_s^2/R^2$ 
are defined with respect to an unperturbed, spherically symmetric WD structure.

In our problem, the WD–IMBH binary is not circular, and multiple
frequencies $\omega_{nlm}$ naturally arise when we consider eccentric orbits. 
In addition, we compute $\boldsymbol{a}_{\rm tidal}$ using the tidal tensor
$C_{ij}$ given in Eq.~\eqref{eq:C_tidal_tensor}  (see also Eqs.~(127)-(134) in \citet{Ishii2005}). 
We then derive the coefficients required to express the
tidal force and the displacement field. For example, to leading order we find 
\begin{equation} \label{Eq: Phi2m detail}
\begin{aligned}
&F_{2m}^{\text{S }}(R,\tau)  =-  \Phi_{2m}/ R =\sum _n F_{n2m}^{\text{S }}(R)\, \mathrm{e}^{-i \omega_{n2m}\tau} \\
&\quad \quad = \frac{G M_{\bullet }}{r ^3}  \left( 1 + \frac{\mathscr{L} }{r ^2} \right)  W_{2m}   \mathrm{e}^{- i m \Psi }  R ,
\end{aligned}
\end{equation} 
where $W_{2\pm2}=\sqrt{3\pi/10}$, $W_{20}=\sqrt{\pi/5}$, and $W_{2\pm1}=0$.
Here, $\Phi_{2m}$ denotes the leading-order coefficient in the spherical-harmonic expansion of the tidal scalar potential, $\Phi_{\rm
tidal}\simeq \sum_m \Phi_{2m} Y_{2m}$, and is related to $C_{ij}$ via the
symmetric-trace-free spherical tensor harmonics \citep{Thorne1980}.  
Our expression for $\Phi_{2m}$ reproduces the classical Newtonian result of
\citet{Press1977}, for which $F_{2\pm 1}^{\text{S }}$ vanishes. This follows
directly from our choice $C_{23}=C_{32}=0$ in
Eq.~\eqref{eq:C_tidal_tensor}. 

One interesting consequence is worth
noting. Since both $r$ and $\Psi$ are functions of the proper time $\tau$,
Eq.~\eqref{Eq: Phi2m detail} implies that the tidal forcing consists of
harmonics associated with two distinct frequencies: the radial orbital frequency and the azimuthal frequency.
This represents a qualitative departure from the Keplerian case, in which only one fundamental frequency exists.

We further note that,
in the outer WD envelope, $\omega_{nlm}^2 \ll N^2, L_l^2$.
In this regime, the WKB approximation applies, and $\chi_{nlm}$, $k_{nlm}^2$, and $V_{nlm}$ reduce to 
\begin{equation}\label{eq: chi k V WKB}
\begin{aligned}
\chi_{nlm}(R)
&\simeq \frac{l(l+1)}{\rho\,\omega_{nlm}^2}, \qquad
k_{nlm}^2(R)
\simeq \frac{l(l+1)}{R^2 \omega_{nlm}^2} N^2, \\
V_{nlm}(R)
&\simeq
\frac{ F_{nlm}^{\text{S }}R}{g } +\frac{F_{nlm}^{\text{R }}-\left ( F_{nlm}^{\text{S }}R \right )^{\prime} }{N^2}.
\end{aligned}
\end{equation}
Here $k_{nlm}^2$ is positive definite, indicating a propagating-wave regime.
Moreover, imposing the physical boundary condition that no waves enter from the stellar surface (or from infinity), the above solution corresponds to a wave that transports energy outward. 
A further simplification follows from our earlier observation that only the quadrupole ($l=2$) component contributes to the tidal tensor.  
In this case, the radial and spheroidal components of $\boldsymbol{a}_{\rm tidal}$ satisfy $F_{nlm}^{\mathrm{R}} \approx (F_{nlm}^{\mathrm{S}} R)'$.
Substituting this relation into   Eq.~\eqref{eq: chi k V WKB} yields
$V_{nlm}\simeq F_{nlm}^{\text{S}}R/g$.

\subsection{Energy and Angular-momentum Fluxes}\label{sec3-3}

Having derived the tidal response of the WD (e.g., $\xi_{nlm}^{\rm R}$), we can compute the energy and angular-momentum transfer from the binary orbit into the star, and thereby infer the corresponding tidal back reaction on the orbital evolution. 
The $z$-component of the time-averaged angular-momentum flux carried by tidally excited gravity waves is given by 
\begin{equation} 
\begin{aligned}
\dot{L}_z(R)
&=\left \langle  \oint  \rho  R^3    \left [  \mathbf{e} _z\cdot \left ( \mathbf{e} _R \times   \boldsymbol{v} \right )   \right ]   \left ( \boldsymbol{v}  \cdot \mathbf{e} _R  \right )   \mathrm{d}\Omega \right \rangle_\tau  \\ 
&\simeq  \frac{\rho  R^3}{2 N}  \sum_{nlm}^{m>0} \frac{ m\, k^{2}_{nlm}R^2}{\left [ l(l+1) \right ]^{3/2} }   \left | \omega _{nlm}  \right |^3 \left |\xi_{nlm}^{\rm R}  \right | ^2    
\end{aligned}
\end{equation} 
\cite{Fuller2012,Vick2017}, where 
$\langle \cdots \rangle_\tau$ denotes averaging over proper time, and
$\boldsymbol{v}=\partial \boldsymbol{\xi}/\partial \tau$
is the fluid three-velocity measured in the FNCs.

The evaluation of $\dot{L}_z(R)$ requires care because the propagation properties of gravity waves vary across the WD interior, owing to the strong spatial variation of the Brunt--Väisälä frequency associated with composition gradients. 
Across a narrow composition-gradient layer bounded by radii $R_a$ and $R_b$, the transition between propagation regimes occurs, where the Brunt--Väisälä frequency varies rapidly, with $N^2(R_a) \gg N^2(R_b)$~\citep{Fuller2012}. 
In this region, the wave propagation changes from partially reflected wave behavior to fully propagating gravity waves.

In the absence of significant local damping or wave reflection, the   angular-momentum flux  varies only weakly across the outer propagation region~\citep{Fuller2012}.  
We therefore evaluate the net flux at the inner edge of the propagation region, $R_a$, i.e.,
$\dot{L}_z(R_\star)\sim \dot{L}_z(R_a)$. 
Assuming that the outgoing gravity waves are ultimately absorbed in the outermost layers of the WD, beyond the propagation region, the flux is transferred to the WD, corresponding to the orbital angular-momentum loss rate, 
$\langle \mathrm{d}L_z/\mathrm{d}\tau \rangle_\tau \sim \dot{L}_z(R_\star)$.

Using the tidal response $\xi^{R}_{nlm}$  obtained in  Eq.~(\ref{eq: Z_second-Eq}), the orbit-averaged rate of change of $L_z$ can be expressed entirely in terms of the tidal forcing and the internal structural properties of the WD.
In practice, the calculation is carried out using dimensionless quantities, 
\begin{equation} \label{Eq: dotLz}
\left \langle  \frac{\mathrm{d}  L_z}{\mathrm{d}  \tau }   \right \rangle_\tau  \simeq   \mathcal{C}_{\star}^{-1} \sum_{nlm}^{m>0}m\, \boldsymbol{\mathcal{F}}_{lm} ( \widetilde{F}_{nlm}^{\text{S }} , \widetilde{\omega}_{nlm}), 
\end{equation}
where $\mathcal{C}_{\star}$  is the compactness of the WD, and
\begin{equation}\label{Eq: F function}
\begin{aligned}
\boldsymbol{\mathcal{F}}_{lm}  ( \widetilde{F}_{nlm}^{\text{S }} , \widetilde{\omega}_{nlm})=& \frac{\hat{f}}{\ \left [ l(l+1) \right ]^{5/2} }\left ( \frac{R_\star}{r_g}  \right ) ^{2l} \times \\
&\widetilde{R}_a^{2l-4} \left |  \widetilde{F} _{nlm}^{\text{S }} ( \widetilde{R}_a) \right | ^2  \left |   \widetilde{\omega} _{nlm}\right |^5.
\end{aligned} 
\end{equation}
Here, quantities with tildes denote the corresponding dimensionless variables.
In particular, $\widetilde{F}_{nlm}^{\text{S }}=F_{nlm}^{\text{S }} (R_a/r_g)^{1-l} (GM_{\bullet}/c^4)$, with $ \widetilde{R}_a=R_a/R_{\star}\sim0.8$, and $   \widetilde{\omega} _{nlm}=\omega_{nlm}/\omega_\star$. 
The dimensionless coefficient  $\hat{f}$ depends only on the internal structure of the WD and must be calibrated using detailed numerical calculations.  
For a typical  $0.6 M_{\odot}$ WD, numerical studies find $\hat{f} \sim 10^{2}$~\cite{Fuller2012, Fuller2013, Vick2017}.

The time-averaged energy dissipation rate can also be expressed in terms of the same function $\boldsymbol{\mathcal{F}}_{lm}$,  
\begin{equation}\label{Eq: dotE}
\left \langle  \frac{\mathrm{d}  E}{\mathrm{d}  \tau }   \right \rangle_\tau \simeq  \frac{\mathcal{C}_{\star}^{1/2}}{q}  \sum_{nlm}^{m\geqslant 0} \left | \omega _{nlm} \right |  \boldsymbol{\mathcal{F}}_{lm} ( \widetilde{F}_{nlm}^{\text{S }} , \widetilde{\omega}_{nlm}).
\end{equation}
Unlike the angular-momentum flux, the energy dissipation receives contributions from 
axisymmetric ($m=0$) modes. This is because axisymmetric perturbations do not exert a net torque, but they do perform net work. 

The dissipation rates derived above are formulated in the FNCs and are referenced to the proper time  $\tau$.
However, orbital evolution is conventionally described in the BLCs.
To relate the two rates, we use  
\begin{equation} 
 \frac{\langle\mathrm{d}E / \mathrm{d}  t\rangle_{t}\ }{\langle\mathrm{d} E / \mathrm{d}  \tau \rangle_{\tau}}  = \frac{\langle\mathrm{d} \tau / \mathrm{d} \lambda\rangle_{\lambda}}{\langle\mathrm{d} t / \mathrm{d} \lambda\rangle_{\lambda}}= \frac{\langle\Sigma\rangle_{\lambda}}{\Upsilon_{t}},
\end{equation}
where $\lambda$ is the Mino time,
$\langle \Sigma \rangle_{\lambda} = \Lambda_r^{-1} \int \Sigma \,\mathrm{d}\lambda$,
and $\Upsilon_t = T_r / \Lambda_r$, with $\Lambda_r$ and $T_r$ denoting the radial periods in Mino time and coordinate time, respectively  \citep{Mino_2003, Fujita_2009}.

The introduction of the Mino time mitigates the slow convergence of Fourier expansions, especially for highly eccentric orbits, such as those appearing in Eq.~\eqref{Eq: Phi2m detail}.
In this work, we implement a two-step frequency-domain method based on the Mino time.
We first expand the tidal factors as Fourier series with respect to the Mino time $\lambda$.
This expansion defines the fundamental frequencies for $1/r$ and $\mathrm{e}^{\pm i\Psi}$ as $\Omega_r = 2\pi/\Lambda_r$ and $\Omega_\Psi = \langle \mathrm{d}\Psi/\mathrm{d}\lambda \rangle_\lambda$, respectively \cite{kerrgeopy2024, Fujita_2009, Dyson_2023}.
We then re-expand the result by expressing the Mino time itself as a Fourier series in the proper time $\tau$, namely,
$
\lambda = \tau /\langle \Sigma \rangle_\lambda + \sum_k C_k \,\mathrm{exp}\left[-ik\Omega_r \tau / \langle \Sigma \rangle_\lambda \right]$.
This re-expansion naturally produces products of finite Fourier series.
We evaluate these products numerically through discrete convolution.
As a result, the method computes the Fourier expansion efficiently  at high harmonic orders.

This improvement allows us to evaluate the travelling-wave tidal response accurately for highly eccentric encounters, and to determine whether the associated orbital energy transfer can compete with the conventional impulsive excitation of the WD $f$-mode during a single pericenter passage.
The outgoing gravity-wave excitation in Eq.~\eqref{Eq: dotE} determines the orbital energy transfer through the Fourier harmonics of the tidal forcing. 
We therefore focus on the evaluation of the forcing kernel $\boldsymbol{\mathcal{F}}_{lm}$ in Eq.~\eqref{Eq: F function}.

We estimate the frequency $\widetilde{\omega}_{nlm}$ and the coefficient $\widetilde{F}^{\rm S}_{nlm}$ by replacing the GR tidal forcing in Eq.~\eqref{Eq: Phi2m detail} with its Keplerian form, since both descriptions agree at leading order in the regime $\mathscr{L}/r^2 \ll 1$. 
We then express this Keplerian forcing in Fourier space to obtain its harmonic decomposition.
The Fourier decomposition reads 
\begin{equation} \label{Eq: Fourier Hansen}
\begin{aligned} 
\frac{\mathrm{e} ^{-im\phi(t)}}{r(t)^3} =\frac{(1-e)^3}{r_p^3} \sum_{K=-\infty }^{+\infty }F_{Km}\mathrm{e}^{-iK\omega _{\rm orb}t} , 
\end{aligned}
\end{equation}
where $F_{Km}$ denotes the Hansen coefficients~\cite{Press1977,Murray1999,Vick2017}. 
These coefficients peak sharply in the high-eccentricity regime $e\simeq0.95$, where the dominant contribution concentrates around a characteristic harmonic number $\bar{K}\simeq 200$ with typical amplitude $\bar{F}_{Km}\sim20$.
We then approximate the full harmonic series in Eq.~\eqref{Eq: dotE} by  the dominant contribution near $\bar{K}$ and replace $F_{Km}$ with its characteristic value $\bar{F}_{Km}$.  
As a result, this approximation yields a tractable scaling for the orbital energy transfer: 
\begin{equation} \label{Eq: deltaEg}
\begin{aligned}
&\frac{\Delta E_{ g }}{M_\star c^2}  =  6\times 10^{-10}
\left ( \!\frac{\hat{f} }{10^2}  \right ) 
\left ( \frac{\bar{F}_{Km}}{20}  \right )^{2}
\left ( \frac{ \bar{K} }{200}  \right )^7\times\\
&
\left ( \frac{ \mathcal{C}_{\star}  }{ 10^{-4} } \right )^{5/2}
\left ( \frac{ q }{0.6/10^5} \right )^{-1}
\left ( \frac{1-e}{0.05}  \right )^{27/2}
\left ( \frac{ k  }{3} \right )^{-27/2}  \!. 
\end{aligned}
\end{equation} 
The dependence on $k$ follows primarily from the $r_p^{-3}\propto k^{-3}$ scaling in Eq.~\eqref{Eq: Fourier Hansen} and from the factor $\left|\widetilde{\omega}_{nlm}\right|^5\propto k^{-15/2}$ in Eq.~\eqref{Eq: F function}.

For comparison, the $f$-mode response mainly arises from impulsive excitation near pericenter rather than from cumulative harmonic driving.
Following Refs.~\cite{Ivanov2004,Yang2018,Wang_2022}, we estimate the energy transfer to the f-mode using the standard impulsive approximation,
\begin{equation} \label{Eq: deltaEf}
\begin{aligned}
&\frac{\Delta E_{ f }}{M_\star c^2} =\frac{16 \sqrt{2}}{15} \widetilde{\omega}_{f}^{3} \widetilde{Q}_{f}^{2} k^{3 / 2} \exp  [-\frac{4 \sqrt{2}}{3} k^{3 / 2} \widetilde{\omega}_{f}  ]\mathcal{C}_{\star}   \\
=&4\times 10^{-10} 
\left ( \frac{ k  }{3} \right )^{3/2}
\exp  [-2.74 \left ( \frac{ k  }{3} \right )^{3/2}]
\left ( \frac{ \mathcal{C}_{\star}  }{ 10^{-4} } \right ) ,   
\end{aligned}
\end{equation}
where $\widetilde{\omega}_{f}=\omega_f/\omega_\star=1.455$ denotes the dimensionless f-mode frequency, and $\widetilde{Q}_{f}=0.5$ denotes the dimensionless tidal overlap integral.
This impulsive excitation decreases exponentially as the pericenter distance increases through the parameter $k$.
We compare Eq.~\eqref{Eq: deltaEg} with Eq.~\eqref{Eq: deltaEf}, and find that the exponential suppression in Eq.~\eqref{Eq: deltaEf} allows the travelling-wave dynamical tide to become comparable to, or to exceed, the $f$-mode energy transfer for $k \gtrsim 3$.

For sufficiently high tidal forcing frequencies in the regime $k \lesssim 3$, the radial wavenumber of gravity waves decreases as $k$ decreases, and the WKB condition gradually loses validity in this limit.
The loss of WKB validity  reduces its reliability in estimating the efficiency of gravity-wave excitation, and then  introduces a systematic overestimate in the inferred gravity-wave energy transfer.
As a consequence, the predicted $\Delta E_g$ can approach values comparable to or even exceed the impulsive $f$-mode energy transfer $\Delta E_f$. 
We therefore treat the predicted energy transfer in Eq.~\eqref{Eq: deltaEg} as an approximate upper envelope in the regime $k \lesssim 3$.
This regime corresponds to $r_p/r_g \lesssim 10$ for a $0.6\,M_{\odot}$ WD orbiting a $10^{5}\,M_{\odot}$ IMBH, with $r_t/r_g \simeq 3.3$ setting the characteristic tidal scale.
The same upper-envelope interpretation remains consistent with the semi-analytic travelling-wave treatments of \citet{Fuller2012} and \citet{Vick2017}, where the approximate $\omega^5$ scaling captures the envelope of the tidal response rather than the detailed structure of individual harmonic contributions.

\section{Results}\label{orbital_evolution_section}

\subsection{Orbital Evolution Around Non-Spinning IMBH}\label{sec:orbitSchw}

Using the prescription for  TD  developed above, we first study the orbital evolution of a WD around a non-spinning IMBH. 
In addition to TD,  GW  radiation is another important driver of the orbital evolution. 
However, the standard PN formalism is inadequate for accurately capturing the dynamics of our system, which resides deep in the strong-gravity regime.
Therefore, we adopt  black hole perturbation theory (BHPT) to
compute energy and angular-momentum loss due to GW radiation for stable, bound,
equatorial Kerr orbits. 
Specifically, for orbits with eccentricities in the range $0 \leq e \leq
0.85$, we rely on the publicly available GW flux data set 
\texttt{FastEMRIWaveforms}~\cite{Chua_2019,Chua_2021,Chua_2021b,Chapman-Bird_2025}. 
For more eccentric trajectories with $0.85 \leq e \leq 0.96$, we perform
direct BH perturbation calculations using
\texttt{pybhpt}~\cite{Nasipak_2022,Nasipak_2024}.

\begin{figure}  % h=here, t=top, b=bottom, p=page
    \centering
    \includegraphics[width= \linewidth]{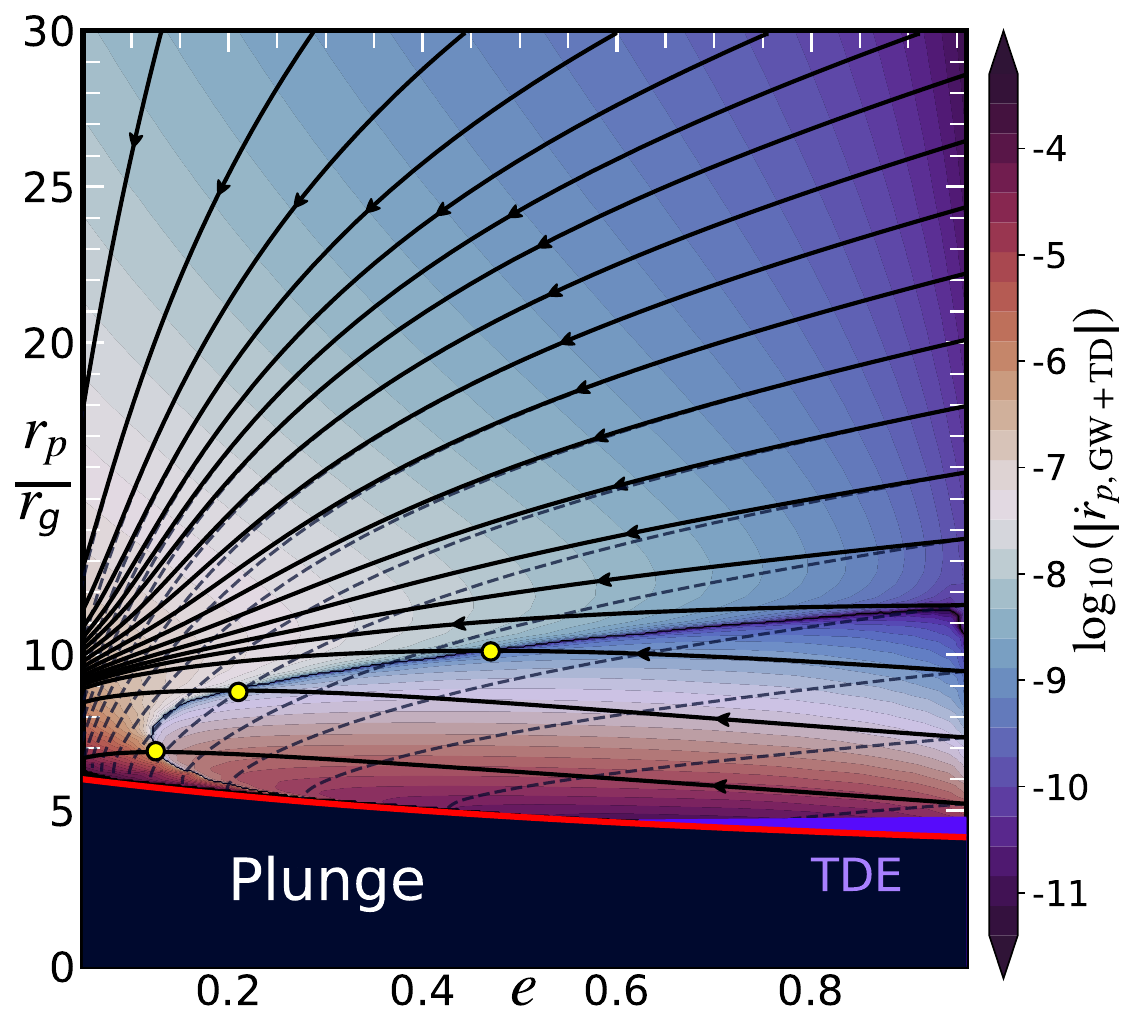}  
    \caption{ 
    Orbital evolution streamlines in the $e-r_{p}$ parameter space for a WD orbiting a non-spinning IMBH under the combined effects of gravitational radiation and TD. 
    Solid curves show orbital trajectories when both mechanisms are included, while dashed curves correspond to evolution driven solely by gravitational radiation. 
    The color map encodes the pericenter evolution rate $\log_{10}\left ( \left | \dot r_{p,\mathrm{GW}}+\dot r_{p,\mathrm{TD}} \right |  \right )  $  under two effects, where $\dot r_{  p}= ({\rm d}r_{  p}/{\rm d}t)/c$. 
    The region below the red curve indicates the plunge region, defined by $r_{p} < r_{\rm plunge} $, where the orbit directly falls into the IMBH event horizon. 
    The purple strip adjacent to the red curve denotes the TDE region, defined by $r_{\mathrm{plunge}} <r_{p} < r_{t}$. 
    Yellow dots mark the locations along the streamlines where the pericenter distance $r_{p}$ reaches a maximum. 
    In this plot, we assume $M_{\star} = 0.6\,M_\odot$ and $M_{\bullet} = 10^{5}\,M_\odot$. 
}
    \label{fig:drpdt_dedt_combined}  
\end{figure} 

FIG.~\ref{fig:drpdt_dedt_combined} shows the evolution in the $e-r_p$ parameter space of a WD orbiting an IMBH under the combined effects of gravitational radiation and TD. 
The color map indicates the pericenter evolution rate $\dot{r}_p$.  
Solid curves denote secular trajectories including both mechanisms, while dashed curves show the evolution driven solely by GW emission. 
The separation between the two sets of curves signals a transition in the dominant dissipation mechanism. 
For sufficiently large pericenter distances ($r_{p} \gtrsim 15\, r_{g}$), the solid and dashed curves nearly overlap, indicating that TD is negligible compared to GW radiation.  
By contrast, at smaller pericenter distances, the two curves deviate, demonstrating that TD becomes dynamically important.

A noteworthy feature is the existence of a region bounded by the condition $\dot r_{p,\mathrm{GW}}+\dot r_{p,\mathrm{TD}}=0$ (e.g., the yellow markers). 
Below this boundary, $r_p$ increases with time.
This behavior is qualitatively distinct from GW-driven binaries, for which $r_p$ decreases monotonically. 
Detecting such a signature in GW or EM observations would provide direct evidence that TD is taking place in WD-IMBH binaries. 
As the binaries evolve past this boundary toward lower eccentricities, GW emission again dominates and drives a decrease in $r_p$.

The origin of this non-monotonic behavior can be understood from the relation $r_p = p/(1+e)$. 
Although TD reduces both the semi-latus rectum $p$ and the eccentricity $e$, the fractional decrease in eccentricity dominates,  i.e., $|\dot p/p| < |\dot e/(1+e)|$. 
As a result, the reduction in $e$ outweighs the shrinkage of  $p$, leading to an increase in $r_p$ despite the overall orbital decay. 

FIG.~\ref{fig:drpdt_dedt_combined} also shows that the inclusion of  TD  can substantially alter the outcome of binary evolution.
First, the solid streamlines show that WD–IMBH binaries are mostly circularized ($e=0$) when TD is included. 
In contrast, the dashed streamlines indicate that, in the absence of TD, many systems remain eccentric by the time they either plunge or undergo tidal disruption.
Second, those binaries that would enter the TDE region if GW radiation is the sole driving mechanism  (e.g., the lowest dashed streamline), now with TD, end up in the plunge region. 
These TD-induced differences would affect the predictions for and interpretations of future GW and EM observations, as we will discuss in Sec.~\ref{sec:6}.

\subsection{Impact of IMBH Spin}

\begin{figure} 
    \centering
    \includegraphics[width= \linewidth]{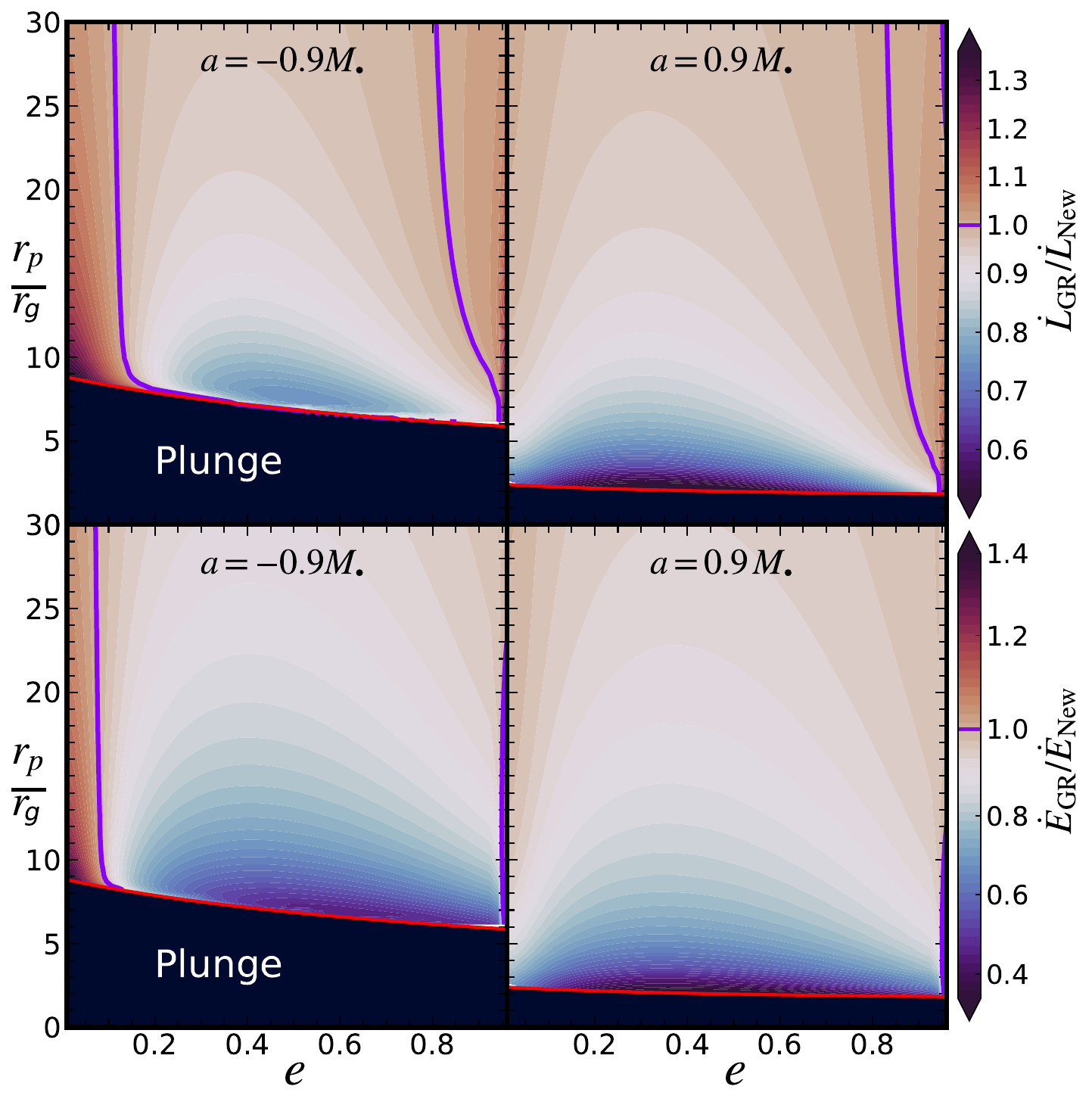} 
    \caption{ 
    Comparison of the TD-induced angular-momentum (upper panels) and energy (lower panels) loss rates computed in the Newtonian and full GR cases. 
    The color maps show the ratios of the rates obtained in the two cases, and the purple curves mark the loci where the ratio equals unity.
    The left and right panels correspond to IMBH spins $a=-0.9M_{\bullet}$ and $a=0.9M_{\bullet}$, respectively.
    All other parameters are the same as those in FIG.~\ref{fig:drpdt_dedt_combined}.
}
\label{fig:DT_GR_Newton_ratio}
\end{figure}

The spin of IMBH shifts the boundary between TDE and plunge orbits, changes the strength of the tidal field, and induces extra rotation of the FNCs, as discussed in Secs.~\ref{sec:2-2} and \ref{sec:3-1}.
FIG.~\ref{fig:DT_GR_Newton_ratio} shows how these effects modify the TD fluxes in a WD–IMBH binary. The fluxes shown here are normalized by their Newtonian counterparts in order to highlight the importance of GR corrections. The Newtonian fluxes are obtained by evaluating the same expressions in Eq.~\eqref{Eq: Phi2m detail} in the Keplerian limit, where relativistic precession is absent.

We find that, over most of the $e-r_p$ parameter space, the orbit-averaged relativistic TD rates are systematically suppressed relative to their Newtonian counterparts.
We emphasize that this suppression does not arise from a weaker instantaneous tidal field. 
On the contrary, near pericenter, the instantaneous relativistic tidal field can exceed the Newtonian one.
The reduction in the net dissipation is a consequence of the loss of temporal coherence in the tidal forcing, induced by the rotation of the FNCs, as discussed in Sec.~\ref{sec:3-1}.

This interpretation is further corroborated by a comparison between the prograde and retrograde cases shown in FIG.~\ref{fig:DT_GR_Newton_ratio}.
For a given pericenter distance with $r_p\lesssim 15\,r_g$, the suppression of the TD rates is more pronounced for retrograde orbits than for prograde ones.
Retrograde orbits, as shown in Sec.~\ref{sec:3-1}, experience a larger effective frame-rotation rate of the FNCs. 
Therefore, the loss of phase coherence in the tidal forcing is stronger, leading to a more prominent suppression of the TD rates.

The trend reverses in the limit of extreme eccentricities, 
e.g., at $e\simeq0$ as well as $e\simeq1$, where we find an enhancement of the TD rates in the GR case.
This enhancement can be understood as follows. 
When $e \simeq 0$, the separation between the WD and the IMBH varies weakly, so that the FNC frame undergoes approximately uniform rotation.
This shifts the tidal forcing to higher frequencies, which, according to Eq.~\eqref{Eq: F function}, generally enhances the stellar response.  
When $e\simeq1$, although we also find a clear enhancement of the TD-induced angular-momentum loss rates in the GR case, the underlying physical mechanism is not yet fully understood. 
A possible contributing factor is that the sharp pericentre passage leads to highly non-uniform tidal forcing,
thereby broadening the effective frequency spectrum, with relativistic effects potentially further amplifying this behavior.

\begin{figure} 
    \centering
    \includegraphics[width=\linewidth]{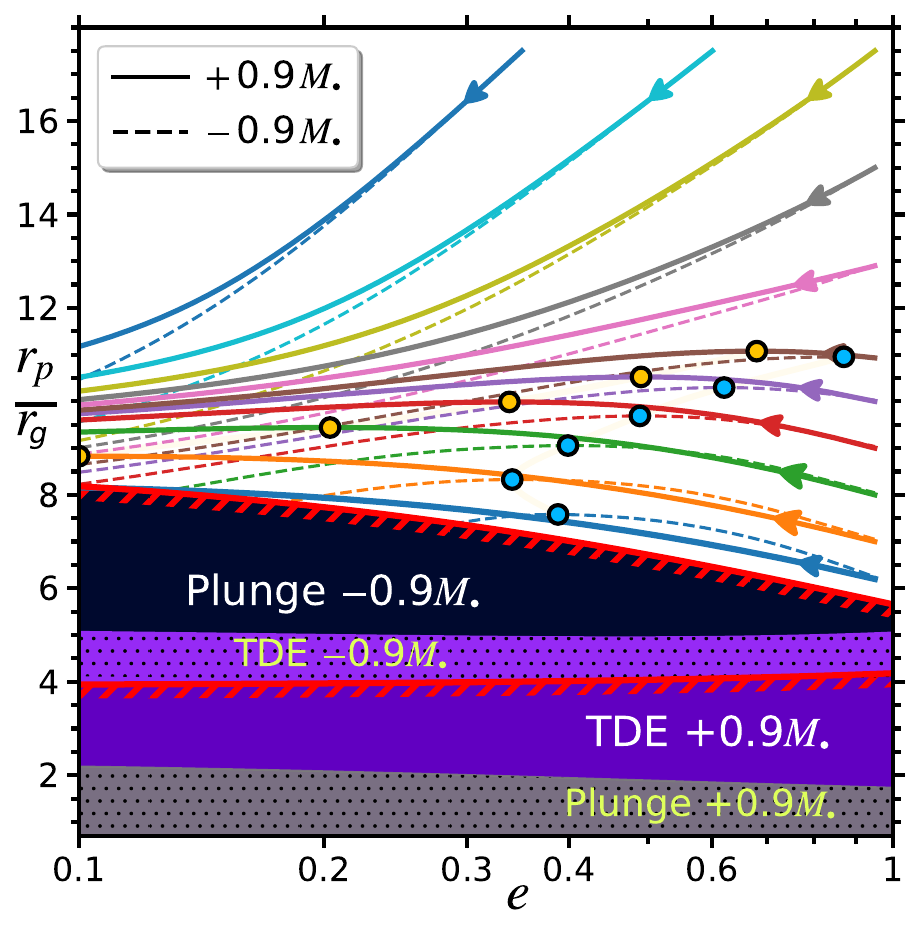} 
    \caption{
    Evolution in the $(r_p,e)$ parameter space for WD-IMBH binaries whose orbital dynamics are driven by GW emission and TD. 
    Solid curves correspond to a spin of $a=0.9 M_{\bullet}$ and the dashed one to $-0.9M_{\bullet}$.
    The yellow and blue dots on these curves mark where $r_p$ reaches maximum, 
    indicating equal contribution of TD and GW in determining the orbital dynamics.
    Shaded regions show where TDEs and plunges will happen (see Section~\ref{sec:2-2} for details). 
    The other parameters are the same as in FIG.~\ref{fig:drpdt_dedt_combined}.
} 
    \label{fig:DT_Trajectory_a_p9_m9}
\end{figure}

Having understood the impact of spin on the TD rates, we can now study the long-term evolution of a WD around a spinning IMBH. 
We notice that the spin-dependent effects enter primarily through the conserved quantity $\mathscr{L} = (L_z - aE)^2$, e.g., in Eq.~\eqref{Eq: Phi2m detail}. 
If $r_p \gg r_g$, we have $L_z/M_{\bullet} \gg E$, and the spin effect will be insignificant.  
Therefore, spin becomes important in the strong-gravity regime, where $r_p$ is not much larger than $r_g$.  
Such a dependence is shown in FIG.~\ref{fig:DT_Trajectory_a_p9_m9}, where we plot representative trajectories for two extreme spin values, $a=\pm 0.9 M_{\bullet}$.  
Large deviation between prograde and retrograde orbits is seen only at $r_p\lesssim 15\,r_g$.

During the spin-dependent evolution, as $e$ decreases and the binary circularizes, a prograde orbit (solid curve) most of the time remains at a larger pericenter distance $r_p$ than a retrograde orbit (dashed curve) of the same $e$.
This general trend is a consequence of two effects operating in different regimes.

First, in the regime where GW radiation dominates the orbital evolution—corresponding to the region above the boundary marked by the yellow (blue) dots in FIG.~\ref{fig:DT_Trajectory_a_p9_m9}—prograde orbits radiate less GW energy and angular momentum than retrograde ones~\cite{Glampedakis2002}.
As a result, their orbital shrinkage (tendency to drift downward) is slower.

Second,   in terms of losing energy via TD,  prograde orbits lose orbital energy more efficiently than retrograde ones, whereas the difference in angular-momentum loss between the two cases is much less prominent.
For example, at a pericenter distance of $r_p \simeq 10\,r_g$ in
FIG.~\ref{fig:DT_GR_Newton_ratio}, the ratio of energy loss
$\dot{E}_{\rm GR}/\dot{E}_{\rm New}$ is $\sim 0.7$ for prograde orbits, but
decreases to $\sim 0.4$ for retrograde ones.
Meanwhile, the corresponding ratio for angular-momentum loss,
$\dot{L}_{\rm GR}/\dot{L}_{\rm New}$ remains nearly constant at
$\sim 0.7$–$0.8$ for both cases.
This different dependence of energy and angular-momentum losses on IMBH spin 
leads to a faster circularization (leftward drift in
FIG.~\ref{fig:DT_Trajectory_a_p9_m9}) of prograde orbits.

Noticeably, in a small region of parameter space with $6\, r_{g} \lesssim r_{p} \lesssim 10\, r_{g}$ and $e \gtrsim 0.4$, the prograde and retrograde orbits behave differently from the general trend discussed above.
In this regime, TD dominates over GW backreaction, and, as shown in FIG.~\ref{fig:DT_GR_Newton_ratio}, retrograde orbits dissipate more orbital energy than prograde ones.
As a result, the pericenters of retrograde orbits increase more significantly during circularization, for the same reason discussed in Sec.~\ref{sec:orbitSchw}.

\section{Gravitational Wave Signatures}
\label{sec:5}

\begin{figure}
    \centering
    \includegraphics[width=\linewidth]{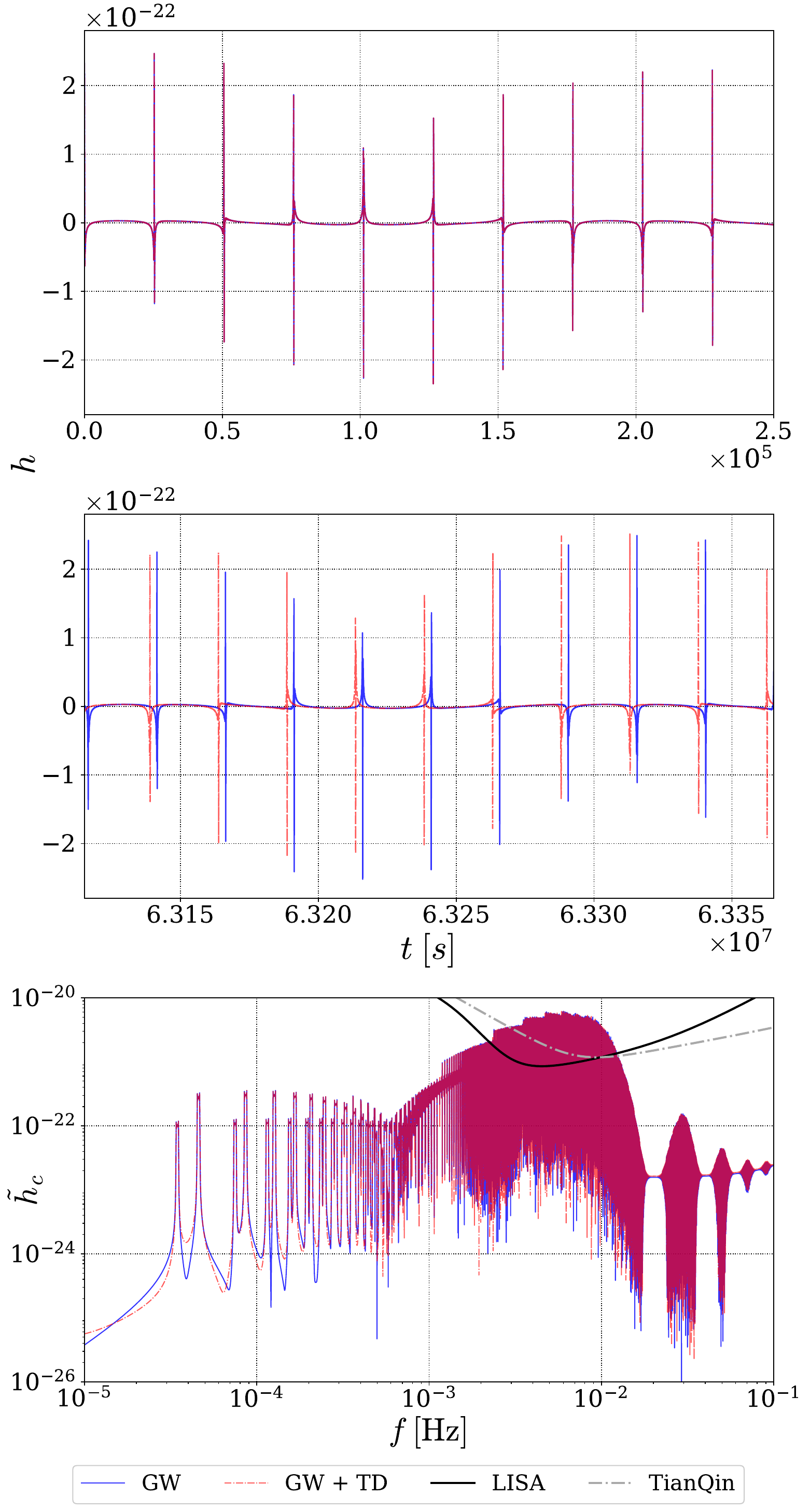}
    \caption{Time-domain waveforms at early (top panel) and late (middle panel) times, and the frequency-domain characteristic strains (bottom panel) 
for binaries driven purely by GW radiation (blue solid curve) versus those binaries
with TD and GW radiation (red dot-dashed curve). Noise curves of 
LISA (black line) \cite{Cornish_2003, Babak_2007, Klein_2016, LISA_2022, LISA_2024} and TianQin (gray dot-dashed line) \cite{TianQin_2021, TianQin_2024} are also shown in the bottom panel. The initial orbital parameters are $( M_\star,M_{\bullet},a, r_{p}, e, x_I)=(0.6M_{\odot},10^5M_{\odot},0.9M_{\bullet},20\,r_{g},0.95, 1)$, and the luminosity distance is $d_{ L}=50\;\mathrm{Mpc}$. The waveforms are computed over the projected 4-year observation time of LISA.}
    \label{TH_GW_comparison}
\end{figure}

\begin{figure}
    \centering
    \includegraphics[width=\linewidth]{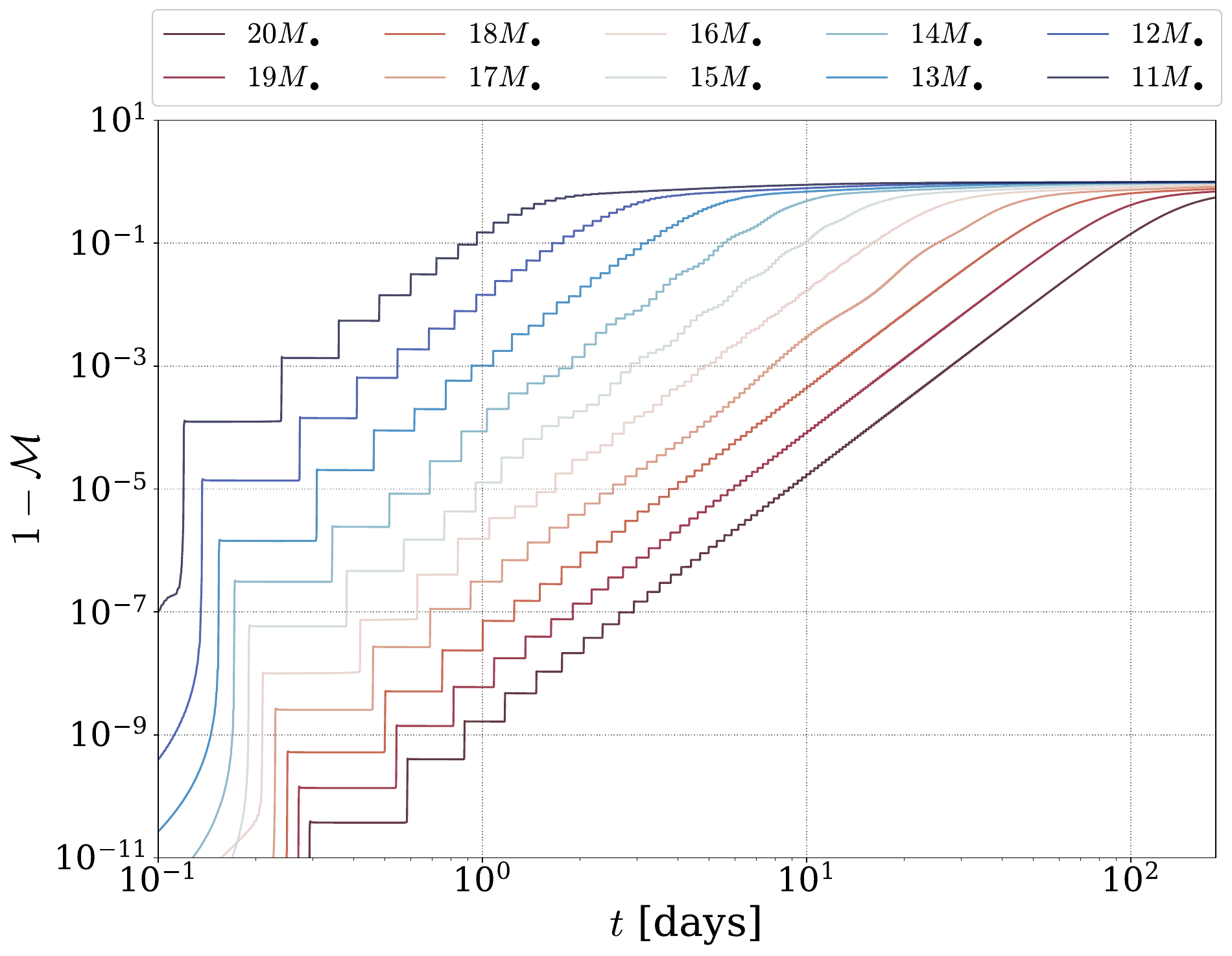}
    \caption{Time evolution of the mismatch between the waveforms with and without TD effects over an observation period of six months. 
    Different lines correspond to different pericenter distances for the WD-IMBH binaries. 
    Here we use the noise curve for LISA~\cite{Cornish_2003, Babak_2007, Klein_2016, LISA_2022, LISA_2024}, and the other initial orbital parameters are the same as in FIG.~\ref{TH_GW_comparison}. }
    \label{mismatch}
\end{figure}

To obtain the GW signals, we first compute the inspiral of a WD-IMBH binary by
evolving the constants of motion using the fluxes. The GW flux is
calculated using the Teukolsky equation~\cite{Teukolsky1972, Teukolsky1973,
Teukolsky1974}, and the TD flux is computed using the method described in
Sec.~\ref{sec3-3}.  From these fluxes, we get a time evolution of
quasi-Keplerian orbital parameters $[p(t), e(t)]$. Since we confine our
study to equatorial orbits, $x_I=\cos\iota=1$, where $\iota$ is the orbital
inclination from the equatorial plane. 

The GW waveform is computed in the frequency domain, in which the $+$- and $\times$-polarizations of the GW strain can be written as
\begin{equation}\label{GW_signal}
    h_{+}+ih_{\times}=\sum_{\ell mnk}\mathcal{A}_{\ell nkm}{^s}S_{\ell m}(\theta,\hat{\omega}_{nkm})e^{i(m\varphi-\hat{\omega}_{nkm}t)}
\end{equation}
\cite{Drasco_2004, Drasco_2006}, where $\mathcal{A}_{\ell nkm}$ is the GW amplitude, ${^s}S_{\ell m}$ is the spin-weighted spheroidal harmonics (for gravitational perturbations we consider $s=-2$), and  $\hat{\omega}_{nkm}=n\hat{\omega}_r+k\hat{\omega}_\vartheta+m\hat{\omega}_\varphi$ are the fundamental frequencies.
In particular, 
$\mathcal{A}_{\ell nkm}$ and $\hat{\omega}_{nkm}$ are functions of $p$ and
$e$, and since these evolve with time, we would in principle need to compute
$\mathcal{A}_{\ell nkm}$ and $\hat{\omega}_{nkm}$ for every time-step in the
orbital evolution. However, in practice, performing such a real-time calculation
of the orbital evolution is far too costly, so we resort to
interpolation schemes similar to those in
\texttt{FastEMRIWaveforms}~\cite{Chua_2019, Chua_2021, Chua_2021b,
Chapman-Bird_2025}. We first compute a grid of values for $\mathcal{A}_{\ell
nkm}$ and $\hat{\omega}_{nkm}$ varying $p$ and $e$ using the \texttt{pybhpt}
software package~\cite{Nasipak_2022, Nasipak_2024}. Then, we create an
interpolant for each $\mathcal{A}_{\ell nkm}$ and $\hat{\omega}_{nkm}$ using
Chebyshev polynomials. Finally, we use the interpolant to extract $\mathcal{A}_{\ell
nkm}$ and $\hat{\omega}_{nkm}$ and, according to Eq.~\eqref{GW_signal}, sum up each mode contribution up to a
tolerance of $\epsilon=10^{-8}$ to derive $h_+$ and $h_\times$.

In FIG.~\ref{TH_GW_comparison}, we compare the GW signal of a WD-IMBH binary
driven by GW radiation and TD versus the signal of a binary driven purely by GW
radiation. The orbital parameters are chosen such that the WD is far from the
TDE radius, so that GW radiation predominates. The corresponding GW radiation
timescale is about $10^3$ yrs.  Despite the secondary role of TD in the orbital
dynamics, the subtle orbital change accumulates coherently over time, so that
in about two years the differences in GW amplitude and phase are clearly
discernible in the time-domain waveform (compare top and middle panels).  

In the bottom panel, we compare the characteristic strains and find that the higher harmonics of the GW signals are in the detection band of LISA and
TianQin. The SNRs for LISA and TianQin are $\rho_{\mathrm{LISA}} = 16$ and
$\rho_{\mathrm{TQ}} = 4.8$, and the matches between the waveforms with and
without TD are $\mathcal{M}_{\mathrm{LISA}}=0.45$ and $\mathcal{M}_{\mathrm{TQ}}=0.53$,
respectively.  Here, $\mathcal{M}$ is defined by
\begin{equation}
    \mathcal{M}(h,\bar{h})=\frac{\langle\tilde{h}|\tilde{\bar{h}}\rangle}{\sqrt{\langle\tilde{h}|\tilde{h}\rangle\langle\tilde{\bar{h}}|\tilde{\bar{h}}\rangle}},
\end{equation}
where $\langle\tilde{h}|\tilde{\bar{h}}\rangle$ denotes the noise-weighted inner product of the Fourier transformed waveforms $\tilde{h}$ and $\tilde{\bar{h}}$~\cite{Owen_1996, Mohanty_1998, Purrer_2020, Pizzati_2022}. These small values of $\mathcal{M}$ indicate that LISA can detect such a binary and
discern the TD effects, and hence distinguish the WD from other compact objects,
 since the threshold for detection is normally $\rho>10$
and the criterion of distinguishing two different waveforms is roughly
$1-\mathcal{M}>1/(2\rho^2)$~\cite{Lindblom_2008}.
However, despite the lower SNR value, TianQin could be useful for 
detecting the TD effects in less distant WD-IMBH binaries with 
smaller pericenter distances~\cite{Torres-Orjuela_2024}. 

We further verify this assertion by plotting the mismatch as a function of time
in FIG.~\ref{mismatch}, and find that the mismatch can reach unity within a
six-month observation period. Observe that as $r_{p}$ decreases, the mismatch
grows more rapidly in time. This is physically consistent since the TD effects
become dominant for pericenter distances close to the IMBH.

\section{Discussion and conclusion} \label{sec:6} 

Motivated by the prospect of observing a WD-IMBH system in both EM and GW
bands, we studied in this work the dynamical evolution of such a system and its
GW signals, with a focus on developing a relativistic model for the response of
the WD to the tidal force of the IMBH.  We first examined the physical scales
and timescales of the system (Sec.~\ref{sec:2}), finding that strong tidal
interaction occurs in the strong-gravity regime when the IMBH mass exceeds
$10^5M_\odot$ (FIG.~\ref{fig:rt_rg_contour_plot}).  Therefore, relativistic
corrections are necessary in modeling the tidal interactions in these systems.
We also found that including the spin of the IMBH could result in a plunge of the
WD into the black hole prior to complete tidal disruption
(FIG.~\ref{fig:rt_rg_contour_plot}), especially for retrograde orbit.  This
result suggests that EM counterparts are more likely to be produced by those WDs on
prograde orbits, which puts strong constraints on the formation channels of
potential multi-messenger sources.

Before the plunge or full tidal disruption, the WD interacts with the IMBH secularly.
the energy and angular-momentum loss from the binary orbit induced by the dynamical (potentially nonlinear) excitation of WD g-modes (Sec.~\ref{sec:3}). 
Our main finding is that TD can be
weaker in our relativistic model than in the previous Newtonian calculations
(FIG.~\ref{fig:DT_GR_Newton_ratio}), despite the fact that the tidal field can be
stronger in our relativistic case. The cause of this suppression of TD 
can be attributed to the loss of temporal phase coherence in tidal
forcing, which is incurred by the rotation of the FNCs (see
Sec.~\ref{sec:3-1}).

\begin{figure}[t]
    \centering
    \includegraphics[width=\linewidth]{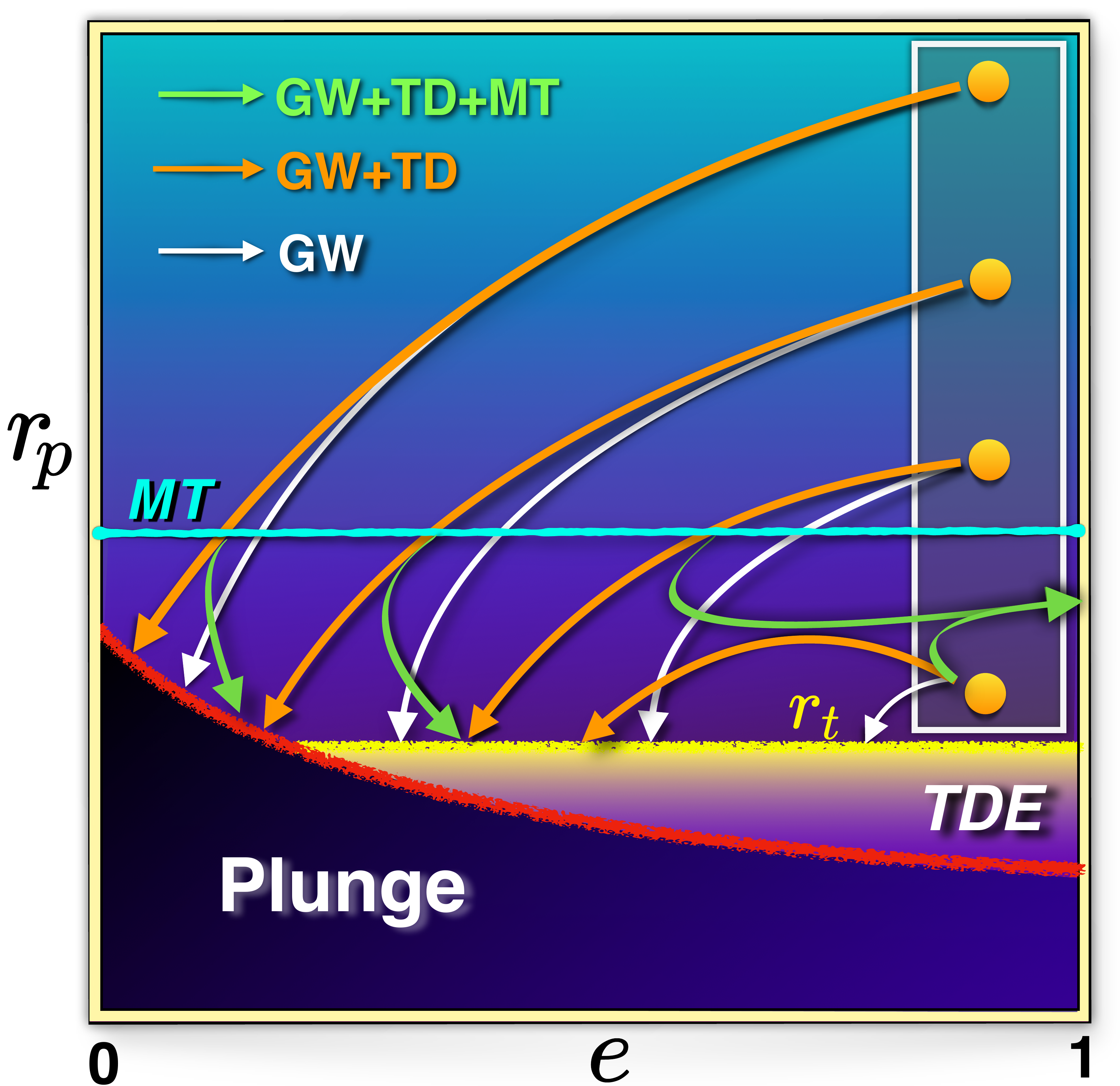} 
    \caption{
    Schematic illustration of WD orbital evolution in the $(r_p,e)$ parameter space.
    Colored trajectories show evolution driven by GW emission alone or by GW combined with TD and MT.
	In addition, the cyan, yellow, and red solid lines indicate the onset of MT, the TDE boundary, and the direct-plunge boundary, respectively.
    }
    \label{fig:WD_rp_e_Trj_show}
\end{figure}

Using the relativistic TD rate, we calculated the secular evolution of the WD-IMBH orbit (FIGs.~\ref{fig:drpdt_dedt_combined} and \ref{fig:DT_Trajectory_a_p9_m9}).  
We find that TD becomes dynamically important once the pericenter distance approaches $\simeq (3-4)\,r_{t}$, corresponding to $r_p \simeq 15\,r_g$ in our fiducial model. 
Moreover, TD leads to more rapid damping of the orbital eccentricity, to the extent that the pericenter distance may even increase over time.
Such an increase in $r_p$ can have important implications for QPEs, as it reduces the cumulative TD rate and delays complete orbital circularization, thereby prolonging the quasi-periodic flaring phase. 
In addition to modifying the flare lifetime, the orbital period evolution may also be affected.
Although TD alone cannot increase the orbital period under mass conservation, its coupling with mass transfer may lead to orbital expansion, potentially accounting for the secular period growth reported in some QPE systems~\cite{Masterson_2025,YangYang2025,Lau2025,Chakraborty_2026}.

An additional effect that can influence this secular evolution arises from the tidal response of the WD spin to TD.
In the present work, we adopt a nonspinning WD configuration, and this assumption removes the coupled evolution between stellar spin and orbital dynamics from the current model.
However, the TD process transfers orbital angular momentum into stellar rotation, and this transfer drives the WD spin toward pseudo-synchronization, a state in which the stellar spin approaches the orbital angular frequency $\omega_p$ near pericenter. 
An order-of-magnitude estimate suggests that spin evolution toward pseudo-synchronization proceeds faster than GW-driven orbital decay, so the system can in principle reach this regime during its evolution. 
This is quantified by comparing the spin synchronization timescale with the GW-driven orbital decay timescale, yielding $t_{\rm synch}/t_{\rm GW} \ll 1$, where $t_{\rm synch}/t_{\rm GW} \simeq \mathcal{C}_{\star}^{5/2}k^{1/2}/q^{2/3}\dot{\mathcal{J}}$, and $I_\star$ is the WD moment of inertia.
As the stellar spin approaches this pseudo-synchronous state, the stellar rotation reduces the effective tidal forcing frequency, and this reduction then suppresses the efficiency of angular momentum transfer \cite{Vick2017}.

However, the stability of pseudo-synchronization remains uncertain in highly eccentric WD–IMBH systems.
The stellar spin evolution in these systems competes with the orbital evolution, and this competition determines whether pseudo-synchronization can persist over long timescales.
Tidal spin-up reduces the effective tidal forcing frequency, and this reduction feeds back on the efficiency of TD.
At the same time, the secular orbital evolution continuously shifts the characteristic tidal frequency as the orbit evolves, and this shift further modifies the spin–orbit coupling.
The interplay between stellar spin evolution and orbital evolution therefore regulates the long-term efficiency of tidal dissipation in these systems.

The TD of orbital energy implies that a
substantial amount of energy is deposited into the hydrogen
envelope of the WD~\cite{Vick2017}. 
According to our model,
the energy dissipation rate can reach 
$10^{32}$--$10^{33}\,\mathrm{erg\,s^{-1}}$. 
If a significant fraction is used to heat the hydrogen envelope, the heating rate already exceeds the typical cooling luminosity of a WD (e.g., $\sim10^{-3}L_{\odot} \sim 10^{30}\,\mathrm{erg\,s^{-1}}$ for a $0.6\,M_{\odot}$ WD) by orders of magnitude.
Therefore, the envelope cannot radiate
the injected energy fast enough, and the temperature will increase.
If the increased temperature triggers thermonuclear runaway at the base of the envelope,  a nova-like
flash would occur. Such a transient EM counterpart 
deserves future investigation.

Finally, we studied the cumulative effect of TD on the GW signal (Sec.~\ref{sec:5}).  We found that even small TD-induced orbital changes, over many pericenter passages, accumulate and produce measurable phase shifts in GW data.  As shown in FIG.~\ref{mismatch},  the mismatch can reach $\sim 0.1$
within months of observation, indicating that TD effects are, in principle, distinguishable in space-based GW data. This suggests that WD–IMBH binaries may be identified not only by their masses but also through tidal imprints in their waveforms, providing a novel probe of WD internal structure.

We find that re-tuning the initial orbital parameters $(a, r_p, e, x_I)$ within the vacuum waveform cannot meaningfully reduce the mismatch with the TD case. 
We performed a targeted analysis restricted to the orbital tangent space spanned by the semi-latus rectum $p$ and eccentricity $e$. 
By computing the noise-weighted inner product between the isolated TD perturbation and the basis vectors $\partial_p h$ and $\partial_e h$, we find that the TD corrections are highly orthogonal to the baseline parameters, yielding overlaps of just $-7\times 10^{-3}$ and $4\times 10^{-4}$, respectively. 
This orthogonality arises because the TD effect introduces a pericentre-localised phase kick, while the baseline parameters control a smooth secular phase evolution of the waveform.
We then invert the localized Fisher matrix to evaluate the corresponding parameter biases, and this inversion shows that the systematic errors induced by neglecting TD remain below the $1\sigma$ statistical uncertainties, with $\Delta p / \sigma_p \approx 0.31$ and $\Delta e / \sigma_e \approx 0.02$. 
Consequently, even an optimal re-tuning of the vacuum parameters cannot dynamically absorb this transient signature, confirming that the mismatch is a robust and distinct consequence of the rapid dephasing caused by TD.

While the present work has focused on equatorial Kerr orbits and $g$-mode
excitation, additional physical processes may further affect dissipation and introduce cycle-to-cycle variability.  
For example, $f$-mode excitation, spin-up of the WD toward a pseudo-synchronous state, and generic orbital inclination are still missing in the current model.  
In addition, MT will occur when $r_p$ approaches the tidal-disruption radius, and previous works have shown that it may increase eccentricity and, in rare cases, render the orbit unbound~\cite{WangDi2024,YangYang2025}.
FIG.~\ref{fig:WD_rp_e_Trj_show} schematically illustrates the possible outcomes under different initial conditions and physical processes. The regime where these outcomes diverge the most corresponds to small pericenter distances, highlighting the importance of general relativistic corrections.  Therefore, a fully relativistic treatment of the TD and MT processes in WD-IMBH systems is essential for making reliable predictions of their multi-messenger signals, and our work may have provided a useful first step.

\section{Acknowledgements}
This work is supported by the National Key Research and Development Program of
China (Grant No. 2021YFC2203002) and the National Natural Science Foundation of China (Grant No. 12473037).  ATO acknowledges support by the National Science Foundation of China (No. W2533010). ATO and LL were supported by Beijing Natural Science Foundation (No. IS25014).

\section{DATA AVAILABILITY}
The data that support the findings of this article are openly available \cite{yang2026data}.

\providecommand{\noopsort}[1]{}\providecommand{\singleletter}[1]{#1}%

\bibliographystyle{unsrtnat}   
\bibliography{main}

\end{document}